\documentclass[useAMS,usenatbib]{mn2e}

\usepackage{natbib,aas_macros,epsfig,graphicx,rotating,lscape,amssymb}

\title[Re-visiting the relations: Galactic thin disc age-velocity dispersion
relation]{Re-visiting the relations:\\Galactic thin disc age-velocity
  dispersion relation}
\author[Seabroke \& Gilmore]{G. M. Seabroke$^{1}$\thanks{E-mail:
gs310@ast.cam.ac.uk}, G. Gilmore$^{1}$\\
$^{1}$Institute of Astronomy, University of Cambridge, UK}

\begin{document}

\date{Accepted . Received ; in original form }

\pagerange{\pageref{firstpage}--\pageref{lastpage}} \pubyear{2007}

\maketitle

\label{firstpage}

\begin{abstract}

The velocity dispersion of stars in the solar neighbourhood thin disc
 increases with time after star formation.  \citet{nordstrom2004} is the most recent observational attempt to constrain the age-velocity dispersion relation.
   They fitted the age-velocity dispersion relations of each Galactic cardinal direction space velocity component, $U$ (towards the Galactic centre), $V$ (in the direction of Galactic rotation) and $W$ (towards the North Galactic Pole), with power laws
 and interpreted these as evidence for continuous heating of the disc in all
 directions throughout its lifetime.  We re-visit these relations with their
 data and use \citet{famaey2005} to show that structure in the local velocity
 distribution function distorts the in-plane ($U$ and $V$) velocity
 distributions away from Gaussian so that a dispersion is not an adequate
 parametrization of their functions.  The age-$\sigma_{W}$ relation can
 however be constrained because the sample is well phase-mixed vertically.   We do not find any local signature of the stellar warp in the Galactic disc.  Vertical disc heating does not saturate at an early stage.  Our new result is that a power law is not required by the data:
    disc heating models that saturate after $\sim$ 4.5 Gyr are equally consistent with observations. 
     
\end{abstract}


\section{Introduction}

The velocity dispersion of the molecular clouds in which star
formation is taking place is very much less, by a factor of three to
five, than is the (one-dimensional) dispersion of old stars near the
Sun. The (set of) processes which change the velocities in any direction of initially slow-moving young
stars into `hotter', higher-dispersion orbits with time are disc `heating' mechanisms and are evidently
fundamental aspects of disc galaxy evolution.  
  
The most recent observational attempt to constrain the age-velocity dispersion relation is by \citet{nordstrom2004}, hereafter N04, using their Geneva-Copenhagen survey of the solar neighbourhood.  This is the largest homogeneous data set available with space velocity components, $U$ (towards the Galactic centre), $V$ (in the direction of Galactic rotation) and $W$ (towards the North Galactic Pole), and isochrone-derived stellar ages.  N04 fitted the age-velocity dispersion relations of each space velocity component (and the total) with continuous smooth power laws: $\sigma \propto \textrm{age}^{k}$, where the scaling (heating) exponent, $k$, is 0.31, 0.34, 0.47 and 0.34 for $\sigma_{U}$, $\sigma_{V}$, $\sigma_{W}$ and $\sigma_{total}$ respectively.  They interpreted these relations as evidence for continuous heating of the disc in all directions throughout its lifetime.  

However, continuous vertical ($\sigma_{W}$) heating is unexpected.
Stars spend most of their 
time near the apocentre of their vertical orbits, furthest away from the Galactic plane.  For
stars to experience continuous vertical heating throughout their lifetimes requires
heating processes that operate away from the plane.  Giant molecular clouds (GMCs) gravitationally scatter stars \citep{spitzer1951,spitzer1953,lacey1984,villumsen1985,hanninen2002} but they exist in the extreme thin disc.  This suggests that vertical heating should not continue throughout the lifetime of the disc but should saturate after some epoch if GMCs are indeed the primary cause.  

There are many heating mechanisms in the literature using non-disc processes to explain continuous vertical heating observations.  \citet{lacey1985}
proposed scattering by massive black holes in the halo, which have been simulated by \citet{fuchs1994} and \citet{hanninen2002}.  \citet{carr1987}
proposed dark clusters of less massive objects could heat stellar discs. However,
neither of these major mechanisms have any direct observational support. 
Minor mergers with dwarf galaxies is another possible heating mechanism \citep{toth1992,walker1996,huang1997}.  Dramatic disc heating can occur when massive satellite galaxies fall on to galactic discs \citep{quinn1993}, which could leave a sharp feature in the age-$\sigma_{W}$ relation but many smaller galaxies accreting could heat a disc smoothly \citep{velazquez1999}.   Other possible
vertical heating mechanisms that may play a minor role are the outer Lindblad
resonance from the Galactic bar
\citep{kalnajs1991,dehnen1999,dehnen2000,fux2001,quillen2003} and evaporating
star clusters \citep{kroupa2002}.  In this paper, we re-visit the
age-$\sigma_{W}$ relation using the same data as N04 to test the uniqueness of
their power law model fit.  We investigate whether the data permit models where the vertical disc heating saturates, thus negating the need to invoke hypothetical or poorly understood components of the Galaxy.

Whether heating is continuous or saturates with time requires separate consideration in the plane ($\sigma_{U}$ and $\sigma_{V}$) and out of the plane ($\sigma_{W}$).   Simulations by \cite{desimone2004} illustrate that transient spiral waves are most likely to be the dominant in-plane heating mechanism in the solar neighbourhood.  Transient spiral waves can lead to gravitational potential fluctuations in the disc that excite random motions of disc stars \citep{barbanis1967,sellwood1984,carlberg1985a,sellwood2002} but only effectively in $U$ and $V$ because they are confined to the thin disc. GMCs also contribute to the in-plane heating but are unlikely to be the dominant heating source \citep{desimone2004}.  Once a star's epicyclic amplitude reaches a size-scale similar to that of the spiral arm radial wavenumbers, the star will cross more than one spiral arm as it radially oscillates making it more difficult for the spiral structure to perturb it \citep{carlberg1985a}.  This means spiral heating is theoretically expected to become much less efficient with time so that its in-plane heating effectively saturates.  

However, whether $\sigma_{U}$ and $\sigma_{V}$ should actually saturate requires simulating the heating from both realistic Galactic spiral structure and GMCs but these calculations are complicated.  Spiral features heat a disc at a rate that depends on the spatial pattern of the potential and on its time variability.  In principle, the photometry of real galaxies can yield the spatial structures of galaxies and temporal information concerning spiral structure can be determined from numerical simulation and dynamical theory.  In practice, the strengths, growth rates and duty cycles of spirals are not well constrained.  Therefore, the relative effectiveness of spiral and GMC heating has to be modelled with an empirical parameter \citep{jenkins1990,jenkins1992}.  Naturally, the empirical parameter was guided by observation and so the models reflected the only observationally determined age-velocity dispersion relation of the time \citep{wielen1977}.  

Like N04,  \citet{wielen1977} also found a continuous power law for
$\sigma_{total}$ (and $\sigma_{W}$) with $k \sim 0.5$ but this was based on
ranking the age of mainly K and M dwarfs using chromospheric emission line
measurements, which is fraught with systematic errors.  \citet{carlberg1985b}
used more reliable isochrone ages and found observational evidence for
saturation in the age-$\sigma_{total}$ relation at ages $>$ 5 Gyr but this was
obtained using tangential velocities and only 255 stars. \citet{quillen2001}
argued for earlier saturation in $\sigma_{U}$, $\sigma_{V}$, $\sigma_{W}$ and
$\sigma_{total}$ between 3 and 10 Gyr.  However, this was based on only 189
\citet{edvardsson1993} stars ($\sim$ 21 stars per logarithmic age bin) and although they used new improved isochrone ages from \citet{ng1998}, their age determination is less reliable than N04.  Also, \citet{stromgren1987}, \citet{freeman1991} and \citet{gomez1997} all argued for a
turnover in the observational relation, indicative of saturation.

The in-plane heating saturation issue is complex because it is not clear if the observations are reliable and the simulations depend on the observations.  In this paper, we argue that this issue is still not observationally resolved even though N04 presented evidence of continuous in-plane heating.  We show their analysis is not conclusive because the interpretation of the in-plane age-velocity
 dispersion relation is complicated by the presence of strong, stellar age dependent, small-scale structure.  
 
\citet{desimone2004} demonstrate that heating by strong transient spiral waves
 can induce strong small-scale structure in the local $U$ and $V$ velocity
 distributions by imparting momentum to stars to put them in specific regions
 of the $UV$-plane, creating streams with a stellar age range of $\sim$ 3 Gyr.
 \citet{sellwood2002} found that spiral waves cause radial migration in the
 galactic disc near their corotation radius, preserving overall angular
 momentum leading to only a small increase in random motions.  This naturally
 explains the `dynamical streams' found in the kinematic survey of local K-M
 giants by \citet{famaey2005}, hereafter F05, which have similar properties to
 moving groups and superclusters but which contain stars spanning a much wider
 age range.  It suggests that the F05 streams are due to an inhomogeneous
 Galactic potential perturbing smooth star formation rather than the streams
 resulting from inhomogeneous star formation in a smooth potential.
 \citet{famaey2007} show that both situations exist in the N04 sample: the
 Hyades-Pleiades stream (see Fig. \ref{fig:uv}) is mainly composed of
 field-like stars but also partly of coeval stars evaporated from the
 primordial Hyades cluster ($\sim$ 85 and 15 per cent of the stream for low-mass stars respectively).  Alternatively, \citet{quillen2005} propose the Hyades-Pleiades and Coma Berenices (or local) streams (middle branch in the top left panel of Fig. \ref{fig:uv})  correspond to nearly closed orbits trapped at the 4:1 inner Lindblad resonance of a two-armed spiral density wave.

\citet{dehnen1998b} used a maximum likelihood algorithm to estimate the
velocity distribution of the N04 stars before their radial velocities were available using only parallaxes and tangential
velocities.  He found the traditionally recognised moving groups as well as a smooth
background distribution, akin to the Schwarzschild ellipsoidal model, as did F05.  The aforementioned streams and the Sirius dynamical stream all share the same region of $UV$ phase-space as the Galactic thin disc background (see Fig. \ref{fig:uv}).  The resulting non-Gaussian $U$ and $V$ velocity distributions pose the question: what parametrization of a complex distribution
 function is adequate to represent a  `dispersion'?

The Gaussianity of the age-$\sigma_{W}$ relation permits it to be re-visited in this paper.  We investigate its derivation from N04 data by checking the stellar warp of the Galactic disc does not contribute to $\sigma_{W}$ and that $W$ is sufficiently well mixed that the presence of streams does not cause $\sigma_{W}$ of the whole sample to deviate radically from its value for the smooth background (as is certainly happening in $\sigma_{U}$ and $\sigma_{V}$).  This is demonstrated by showing that dynamical streams do not affect $\sigma_{W}$ by deriving the age-$\sigma_{W}$ relation with and without the Hercules stream. 

The Hercules stream is the only stream in the N04 sample that does not share the same region of $UV$ phase-space as the Galactic thin disc background, allowing it to be unambiguously defined in $UV$ phase-space and removed in $W$ without also excluding a large region of the background ellipsoid (as would happen in removing one of the other streams).  The Hercules stream has been speculated to be a signature of the
Galactic bar.  Whether Hercules stream stars are chaotically,
gravitationally scattered from the inner Galaxy by the Galactic bar
\citep{fux2001} or they are more local stars coupled to the outer
Lindblad resonance \citep{dehnen1999,dehnen2000} is still uncertain.
\citet{quillen2003} confirms that the Hercules stream remains a strong feature of the local velocity distribution function when the effect of spiral structure is added to the Galactic bar.
\citet{bensby2007} investigated the abundance and age structure of the
Hercules stream.  They found it contains a mixture of young and old stars, all with $UV$ kinematics more characteristic of the thick disc
than the thin disc.  

Section \ref{s:data} describes the N04 data used in this paper: survey design
(Section \ref{s:design}), space velocities (Section \ref{s:uvw}) and discusses
the N04 age derivation (Section \ref{s:ages}).  Our analysis is presented in
Section \ref{s:analysis}, where we reproduce the N04 age-velocity dispersion
relation age bins (Section \ref{s:agebins}) and physically interpret them by
qualitatively comparing them to the F05 velocity distributions (Sections
\ref{s:kine} and \ref{s:phase}).  Section \ref{s:warp} confirms the stellar
warp of the Galactic disc does not contribute to $\sigma_{W}$.  We re-visit
the age-$\sigma_{W}$ relation using the same age binning as N04 in Section \ref{s:age25} and with our own higher resolution age binning in Section \ref{s:age1}, before discussing our model fitting results in Section \ref{s:discussion}.

\section{Data}
\label{s:data}

\subsection{Survey design}
\label{s:design}

 N04 presented a complete, all-sky, magnitude-limited, and
 kinematically unbiased sample of 16,682 nearby F and G dwarf stars.  F
 and G dwarf stars are good probes of Galactic evolution because they
 are numerous and long-lived.  Their radial velocities can be measured
 accurately and their ages can be estimated by comparison with stellar
 evolution models, at least for the more evolved stars.  Ages are
 crucial to place the kinematic properties of stars in an evolutionary
 context.  The following section summarizes the properties of N04 most
 pertinent to re-visiting the age-velocity dispersion relation.

 The N04 observational input catalogue was selected from a compilation
 of catalogues available in the literature with Str\"omgren $uvby\beta$
 photometry of nearby F and G stars, mainly from the surveys by
 \citet{olsen1983, olsen1993, olsen1994a, olsen1994b}. The final sample
 is volume complete to $\approx$ 40 pc. It is magnitude complete to $V
 \le 7.7$ for the bluest stars (slightly fainter for the reddest G
 stars) and has a $V$ cut-off magnitude of $\approx$ 8.9 (and 9.9 for
 the reddest G stars).

\subsection{Space velocities}
\label{s:uvw}

 The input catalogue was observed using the CORrelation RAdial
 VELocities (CORAVEL) photoelectric cross-correlation spectrometers
 \citep{baranne1979, mayor1985}.  Two CORAVELs cover the entire sky
 between them: one is on the Swiss 1-m telescope at Observatoire de
 Haute-Provence, France, and the other on the Danish 1.5-m telescope at
 ESO, La Silla.  Their fixed, late-type cross-correlation template
 spectra match the spectra of the majority of the input catalogue
 stars.  The multi-epoch radial velocities (generally two or more
 spectroscopic observations) have a typical mean error of 0.5 km
 s$^{-1}$ or less.

 The vast majority of the N04 stars have proper motions in the {\it
 Tycho-2} catalogue \citep{hog2000}.  This catalogue was constructed by
 combining the {\it Tycho} star-mapper measurements of the {\it
 Hipparcos} satellite with the Astrographic Catalogue based on
 measurements in the Carte du Ciel and other ground-based catalogues. A
 small number of stars have only one measurement, either from {\it
 Hipparcos} or {\it Tycho}. The typical mean error in the total proper
 motion vector is 1.8 mas yr$^{-1}$.

 The primary source of distance for N04 stars is {\it Hipparcos}
 trigonometric parallax ($\pi$, \citealt{esa1997}), from which absolute
 magnitude ($M_{V}$) can be derived. This is adopted if its relative
 error ($\sigma_{\pi}/\pi$) is accurate to 13 per cent or better,
 otherwise the photometric distance calibrations for F and G dwarfs by
 \citet{crawford1975} and \citet{olsen1984} are used, with an
 uncertainty of only 13 per cent.  Distances are not provided for stars
 with unreliable {\it Hipparcos} parallaxes ($>$ 13 per cent) and when
 photometric distances cannot be calculated.  This occurs when the star
 is missing the necessary photometry and/or it falls outside the
 photometric calibrations.  The absence of a distance estimate or
 radial velocity measurement reduces the size of the N04 catalogue with
 all full six dimensional phase-space information to 13,240 stars.   The
 space velocity components ($U$, $V$, $W$) are computed for all the
 stars with (mean) radial velocities, proper motions and distances.  Space velocities in the VizieR catalogue accompanying N04 \citep{gc_cat2004} are
provided to the nearest km s$^{-1}$.  The average error in each component is 1.5 km s$^{-1}$.

\subsection{Ages}
\label{s:ages}

\subsubsection{N04 age derivation}

N04 derive isochrone ages for their stars.  Ages are derived by using the photometrically measured effective temperatures ($T_{\mathrm{eff}}$), $V$ magnitudes and metallicity ([Fe/H]) to place stars in the Hertzsprung-Russell diagram and reading off the age of the stars by interpolation between theoretically computed isochrones.  $T_{\mathrm{eff}}$, $V$ and [Fe/H] all require extinction correction.  The Str\"omgren $\beta$ photometry provides interstellar reddening,
 $E(b - y)$, for F stars via the intrinsic colour calibration by
 \cite{olsen1988}.  This has been applied in the photometric
 temperature and distance determination if $E(b - y) \ge 0.02$ and the
 distance is above 40 pc, otherwise the stars are assumed to be
 unreddened.  $T_{\mathrm{eff}}$ is 
 calculated from the reddening-corrected $b - y$, $c_{1}$ and $m_{1}$
 indices and the calibration of \citet{alonso1996}, which is based on
 the infrared flux method.  
 
 The majority of N04 metallicity values are derived using the F or G star
 calibrations of \citet{schuster1989}.  Due to the very few spectroscopic
 calibrators available at the time, the \citet{schuster1989} calibration
 yields systematic errors for the reddest G and K dwarfs ($b - y > 0.46$).  To
 remedy this, N04 performed a new fit of the $uvby$ indices to [Fe/H] derived
 from high-resolution spectra for 72 dwarfs ($0.44 \le b - y \le 0.59$), using
 the same terms as the \citet{schuster1989} G star calibration.  When valid,
 N04 adopted the \citet{edvardsson1993} $\beta$ and $m_{1}$ calibration for
 stars with high $T_{\mathrm{eff}}$ and low gravities outside the
 \citet{schuster1989} calibrations.  For stars outside both calibrations, N04
 used other spectroscopic sources to derive a new relation using the same
 terms as the \citet{schuster1989} F star calibration.  The typical [Fe/H]
 uncertainty is $\sim$ 0.1 dex.

Individual stellar ages are determined by a Bayesian estimation method.  For every point in a dense grid of interpolated Padova isochrones \citep{girardi2000,salasnich2000} and given a star's nominal position in the three dimensional Hertzsprung-Russell `cube' defined by log $T_{\mathrm{eff}}$, $M_{V}$ and [Fe/H], N04 compute the probability that the star is actually located at every point.  This procedure takes into account observational errors \citep{jorgensen2005}.  Integrating over all the grid points gives the global likelihood distribution for the possible ages of the star.  N04 account for statistical biases in the integration.  The resulting probability distribution function (`G-function') is normalised to unity at maximum (see their fig. 13).  A stable age estimate without significant bias is given by the maximum of the smoothed G-function.  Upper ($\sigma_{\mathrm{age}}^{\mathrm{high}}$) and lower
($\sigma_{\mathrm{age}}^{\mathrm{low}}$) 1$\sigma$ confidence limits 
are the points where the G-function reaches a value of 0.6.  

Isochrone ages can only be
 determined in practice for stars that have evolved significantly away
 from the Zero Age Main Sequence. Therefore, unevolved
 (i.e. relatively young) stars in the N04 catalogue do not have
 reliable isochrone ages.  This does not affect this study as we are
 concerned with the functional form of the age-velocity dispersion
 relation at the evolved end (i.e. old stars) and do not require a complete or unbiased age distribution function.

\subsubsection{Age comparisons between N04 and other data sets}

\citet{holmberg2006} substantially improved
 temperature and metallicity calibrations of the photometry used by
 N04, evaluated their effects on the computed ages and performed
 extensive numerical simulations to verify the robustness of the
 derived relations.  They found good agreement between their new
 ages and those of N04, supporting the N04 results.  
 
 However, \citet{haywood2002} demonstrates that the original calibration by
 \citet{schuster1989} is subject to systematic errors, mainly due to incorrect
 placement of the Hyades reference sequence.  Using the same formalism as
 \citet{schuster1989}, \citet{haywood2002} derived an alternative calibration,
 which is in good agreement with spectroscopic data sets.  \citet{reid2007}
 calculated the [Fe/H] difference between \citet{haywood2002} and N04 and
 found systematic differences between -0.081 and -0.015 dex for different
 colour ranges. \citet{reid2007} also compared the photometrically-derived
 [Fe/H] of N04 and \citet{haywood2002} with high-resolution echelle
 spectroscopically-derived [Fe/H] obtained by the Keck, Lick and AAT planet
 search programs \citep{valenti2005}.  They found that while the dispersions
 were similar ($\sim$ 0.1 dex), the difference between \citet{valenti2005} and \citet{haywood2002} (-0.041 to +0.045 dex) is smaller than between \citet{valenti2005} and N04 (0.015 to 0.090 dex).
 
 Since the N04 ages are derived from isochrone fitting, systematic [Fe/H] errors translate to systematic age offsets.  As a recalculation of the N04 ages to investigate the effect of systematic [Fe/H] errors is unnecessary for this paper, we proceed with the assumption that any age offset will have a negligible impact on our statistical use of age in this study.  In addition, we are only using single stars with well-defined ages and age errors, so we assume the N04 age errors are the appropriate errors.
 
 \begin{figure*}
\begin{minipage}{170mm}
\centering
\psfig{figure=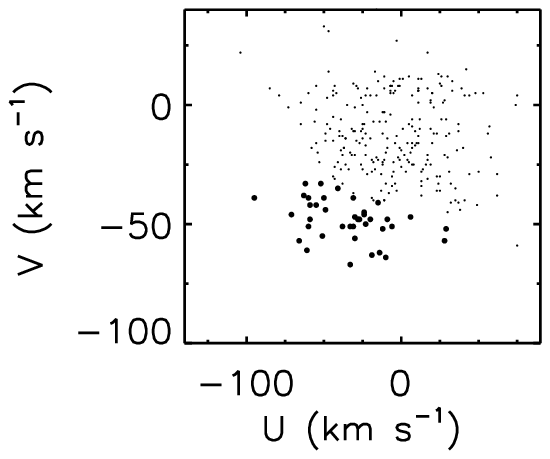} 
\psfig{figure=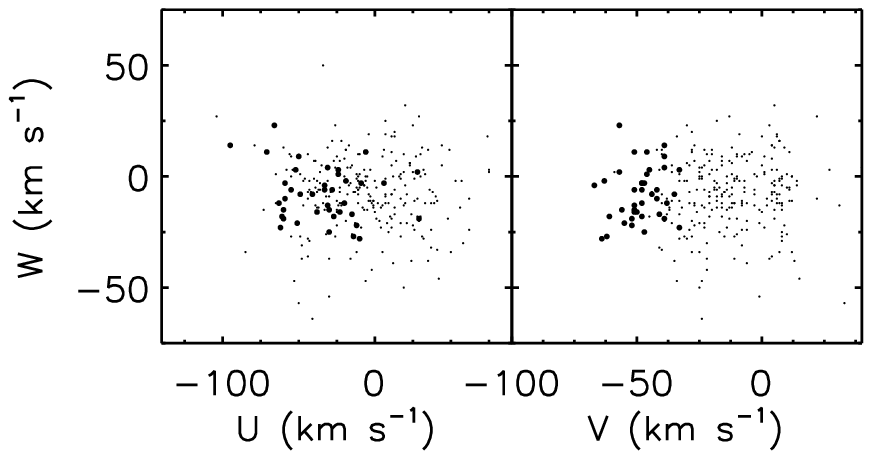}
\caption{$U-V$ (left), $U-W$ (middle) and $V-W$ (right) space velocity diagrams of
  the 278 N04 single stars with relative age errors $<$ 25 per cent in the 2.8
  $<$ age $\le$ 4.6 Gyr age bin.  We have assigned 39 N04 stars in this age bin to the
        Hercules stream (filled circles) using the Hercules stream
        $UV$ phase-space (defined by F05, see Fig. \ref{fig:uv}) and the Hercules stream
        $W$ phase-space (defined by F05, see Figs. \ref{fig:uw} and \ref{fig:vw}).}
\label{fig:uvw25age}
\end{minipage}
\end{figure*}

 \begin{figure*}
\begin{minipage}{170mm}
\centering
 \psfig{figure=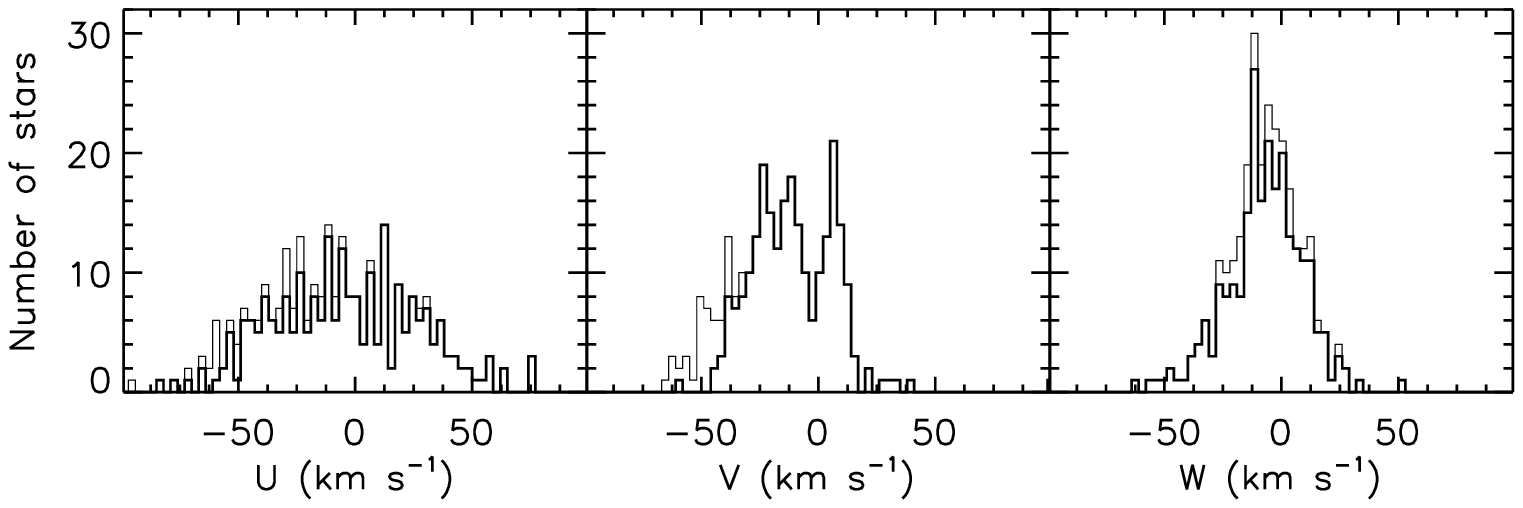}
\caption{$U$ (left), $V$ (middle) and $W$ (right) velocity distributions of
 the N04 single stars with relative age errors $<$ 25 per cent in the 2.8
  $<$ age $\le$ 4.6 Gyr age bin (thin
  line) and a subsample of these stars that excludes the Hercules stream
  (thick line), as defined in
  Figs. \ref{fig:uv}, \ref{fig:uw} and  \ref{fig:vw}.} 
\label{fig:uvw25hist}
\end{minipage}
\end{figure*}

\begin{figure*}
\centering
\begin{minipage}{170mm}
        \psfig{figure=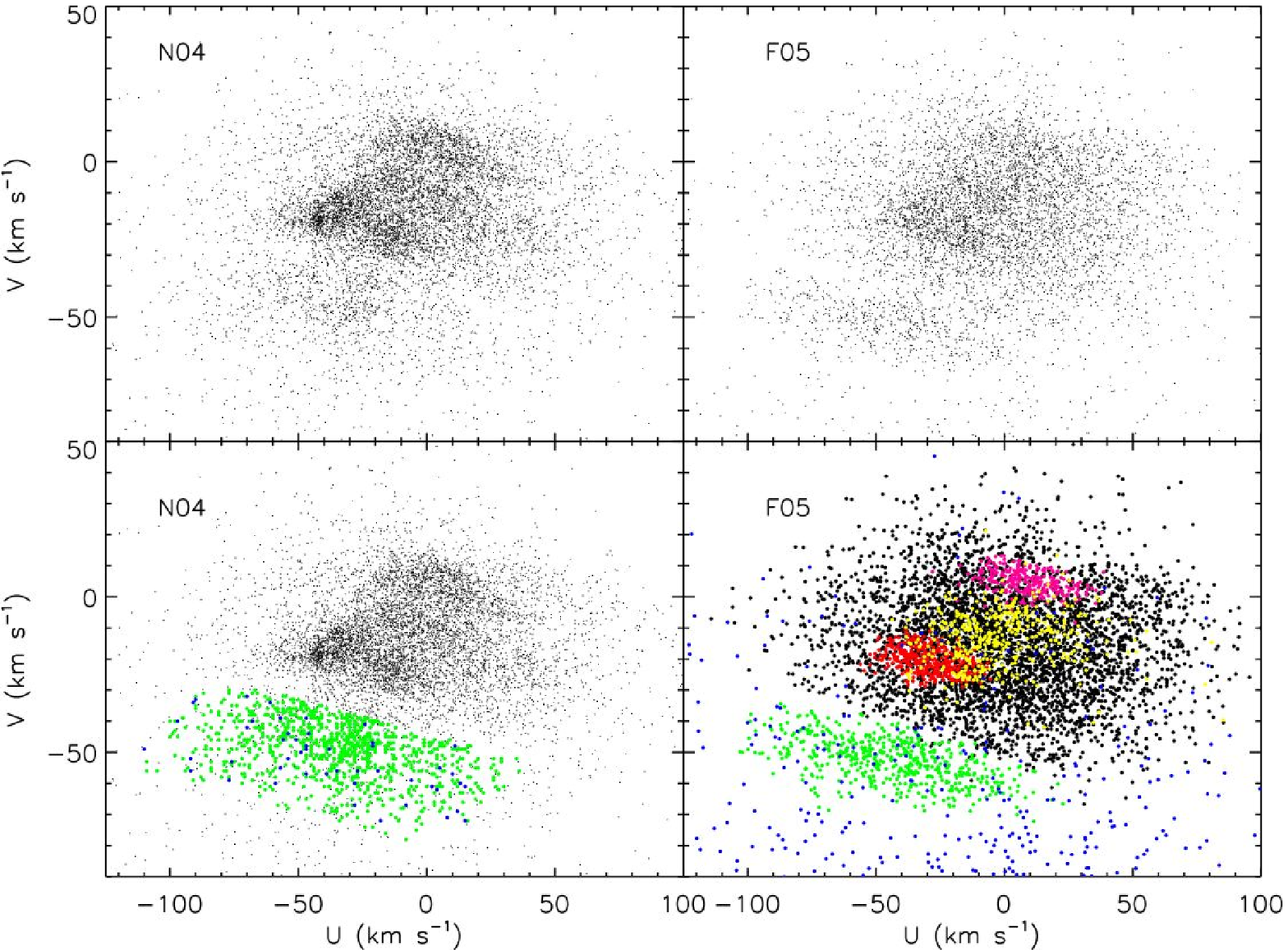,width=1\columnwidth}
\caption{$U$ - $V$ diagrams of the 8,589 N04 single
        F-G dwarfs with space velocities (left column) and the 6,030 F05 K-M
        giants with space velocities, including spectroscopic binaries with
        centre-of-mass radial velocities (right column).  The overdensity at
        ($U$,$V$) = (-45,-20) km s$^{-1}$ in the left column is the
  Hyades open cluster (only in the N04 sample).  In the bottom left
        panel, the N04 stars are colour-coded according to our assignment of N04 stars to the
        Hercules stream (green) and high-velocity stars (blue) in the Hercules stream
        $UV$ phase-space (defined by F05, bottom right panel) but outside the Hercules stream
        $W$ phase-space (see Figs. \ref{fig:uw} and
\ref{fig:vw}).  In the bottom right
        panel, the F05 stars are colour-coded according to their maximum
        likelihood base group
        assignment to known kinematic features of the solar neighbourhood: smooth
  background (black), Sirius moving group (magenta), young
  kinematics (yellow), Hyades-Pleiades supercluster (red),
  Hercules stream (green) and high-velocity stars (blue).}  
                \label{fig:uv}
                \end{minipage}
\end{figure*}

\begin{figure*}
\centering
\begin{minipage}{170mm}
        \psfig{figure=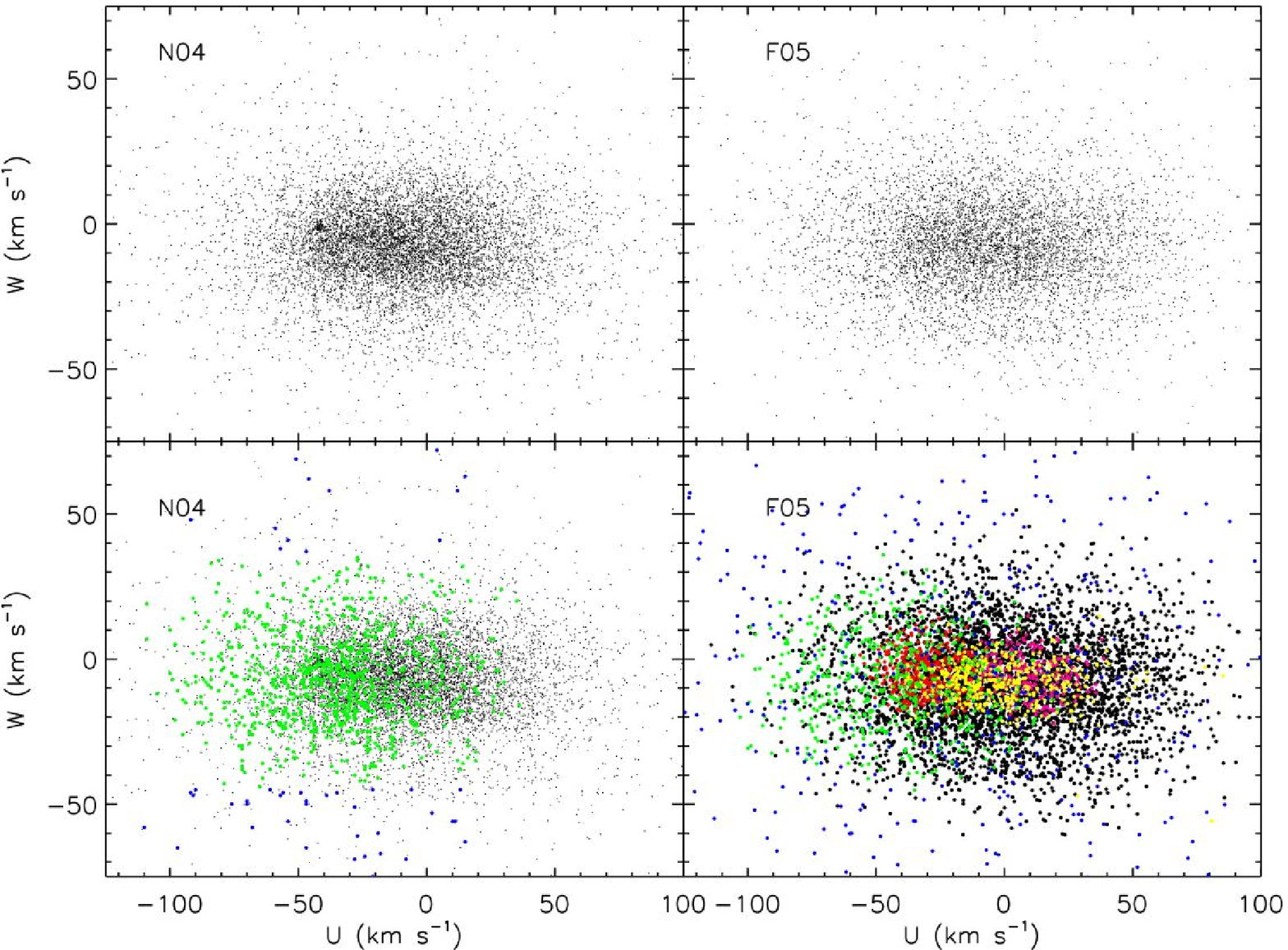,width=1\columnwidth}
\caption{$U$ - $W$ diagrams of the 8,589 N04 single
        F-G dwarfs with space velocities (left column) and the 6,030 F05 K-M
        giants with space velocities, including spectroscopic binaries with
        centre-of-mass radial velocities (right column).  The overdensity at
        ($U$,$W$) = (-45,0) km s$^{-1}$ in the left column is the
  Hyades open cluster (only in the N04 sample).  In the bottom left
        panel, the N04 stars are colour-coded according to our assignment of N04 stars to the
        Hercules stream (green) and high-velocity stars (blue) in the Hercules stream
        $UV$ phase-space (defined by F05, see Fig. \ref{fig:uv}) but outside the Hercules stream
        $W$ phase-space (defined by F05, bottom right panel).  In the bottom right
        panel, the F05 stars are colour-coded according to their maximum
        likelihood base group
        assignment to known kinematic features of the solar neighbourhood: smooth
  background (black), Sirius moving group (magenta), young
  kinematics (yellow), Hyades-Pleiades supercluster (red),
  Hercules stream (green) and high-velocity stars (blue).}  
                \label{fig:uw}
                \end{minipage}
\end{figure*}

\begin{figure*}
\centering
\begin{minipage}{170mm}
        \psfig{figure=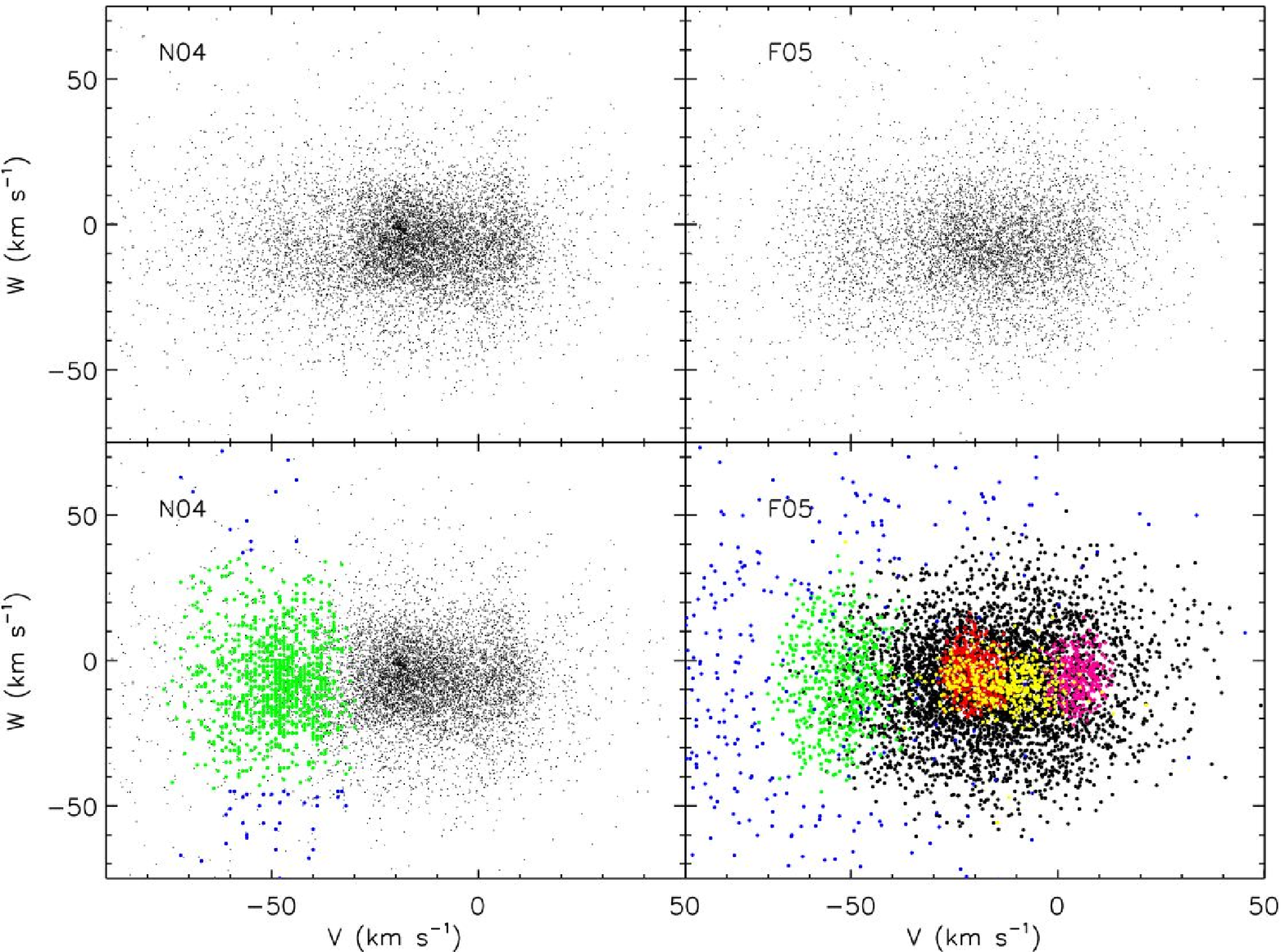,width=1\columnwidth}
\caption{$V$ - $W$ diagrams of the 8,589 N04 single
        F-G dwarfs with space velocities (left column) and the 6,030 F05 K-M
        giants with space velocities, including spectroscopic binaries with
        centre-of-mass radial velocities (right column).  The overdensity at
        ($V$,$W$) = (-20,0) km s$^{-1}$ in the left column is the
  Hyades open cluster (only in the N04 sample).  In the bottom left
        panel, the N04 stars are colour-coded according to our assignment of N04 stars to the
        Hercules stream (green) and high-velocity stars (blue) in the Hercules stream
        $UV$ phase-space (defined by F05, see Fig. \ref{fig:uv}) but outside the Hercules stream
        $W$ phase-space (defined by F05, bottom right panel).  In the bottom right
        panel, the F05 stars are colour-coded according to their maximum
        likelihood base group
        assignment to known kinematic features of the solar neighbourhood: smooth
  background (black), Sirius moving group (magenta), young
  kinematics (yellow), Hyades-Pleiades supercluster (red),
  Hercules stream (green) and high-velocity stars (blue).}  
                \label{fig:vw}
                \end{minipage}
\end{figure*}

\begin{figure*}
\begin{minipage}{170mm}
\centering
\psfig{figure=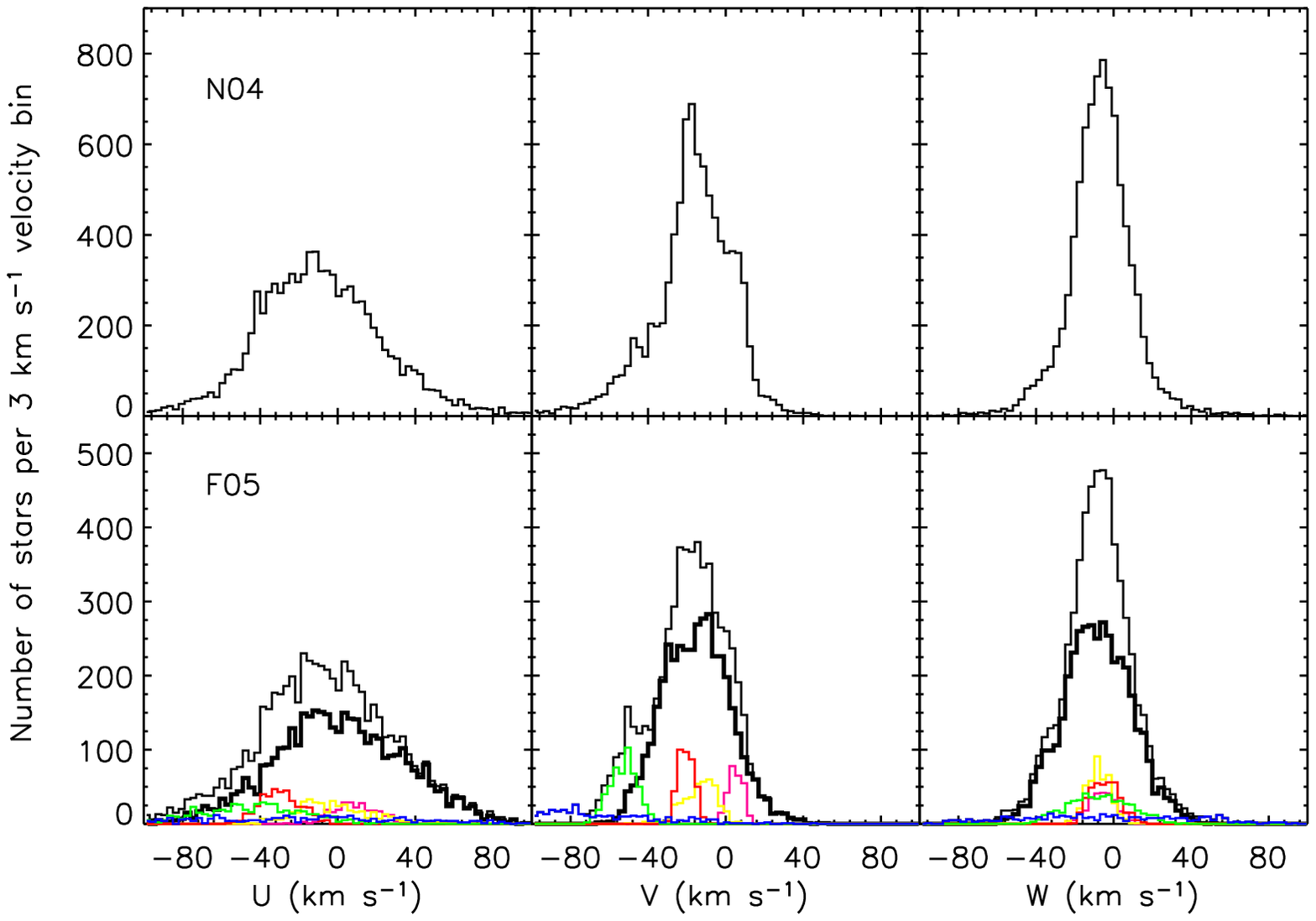,width=1\columnwidth}
  \caption{$U$ (left), $V$ (middle) and $W$ (right) velocity distributions of the 8,589 N04 single
        F-G dwarf stars with space velocities  (top row, thin black lines) and the 6,030 F05 K-M giants
        with space velocities (bottom row, thin black lines), including spectroscopic binaries with
        centre-of-mass radial velocities, colour-coded according to their maximum
        likelihood base group
        assignment to known kinematic features of the solar neighbourhood: smooth
  background (thick black lines), Sirius moving group (magenta), young
  kinematics (yellow), Hyades-Pleiades supercluster (red),
  Hercules stream (green) and high-velocity stars (blue).}
     \label{fig:uvwgcf}
\end{minipage}
\end{figure*}

\section{Analysis}
\label{s:analysis}

\subsection{Reproducing the N04 age-velocity dispersion relation age bins}
\label{s:agebins}

N04 caution that many stars in their sample are binary or multiple
  systems, for which the derived metallicities and thus ages will be
  unreliable.   They provide a catalogue flag ($f_{b}$) that identifies confirmed
or suspected binaries, where the information can come from one or
several sources such as photometry, radial velocity or astrometry.  N04 used only single stars (as defined by a null $f_{b}$ catalogue entry, 8,589 stars) 
with relative age errors $<$ 25 per
cent (2,852 stars) to derive their age-velocity disperson relations in their fig. 31.  Our attempt to reproduce this sample by rejecting stars with $f_{b}$ catalogue entries to define single stars and selecting stars with ($\sigma_{\mathrm{age}}^{\mathrm{high}}$ - age)/age $\le$ 0.25 and (age
- $\sigma_{\mathrm{age}}^{\mathrm{low}}$)/age $\le$ 0.25 yields a smaller
  sample (2,801 stars).  The missing 51 stars are presumably due to N04 using
  more precise ages than their  catalogue ages, which have been rounded up to
  the nearest 0.1 Gyr.

N04 split their sample of single stars with relative age errors $<$ 25 per
cent into ten age bins of equal numbers.  Due to the presumably lower age
resolution in their catalogue, we are unable to make the numbers of stars in
each of our age bins exactly the same.  This means that the boundaries of our
age bins are only similar to those used by N04, rather than being exactly the
same.  Unlike N04, we plot the space velocity diagrams
(Fig. \ref{fig:uvw25age}) and histograms
(Fig. \ref{fig:uvw25hist}) of a representative age bin (plots of all the age
bins are in the Appendix).  The 3 km s$^{-1}$ velocity bin sizes in
Fig. \ref{fig:uvw25hist} (and Fig. \ref{fig:appendixuvw25hist}) are chosen to
be twice the average error in each component of the N04 space velocities so
that substructure can be resolved (F05 found the Hyades-Pleiades and Sirius
dynamical streams have $\sigma_{V} \approx$ 5 km s$^{-1}$).  From the apparent lack of substructure in Fig. \ref{fig:uvw25age} (and
Fig. \ref{fig:appendixuvw25age}), it is not clear in Fig. \ref{fig:uvw25hist} (and Fig. \ref{fig:appendixuvw25hist}) whether the
peaks in the histograms are Poisson noise, due to the relatively small number
of stars in each bin at this velocity resolution, or substructure.  Decomposition of the N04 sample into its constituent kinematic groups, like F05, is beyond the scope of this paper.  Instead, we qualitatively compare the N04 and F05 space velocity diagrams and histograms to assess whether we expect complex substructure in all the age bins in Figs. \ref{fig:appendixuvw25age} and \ref{fig:appendixuvw25hist} or whether the peaks are noise such that the distribution functions can be approximately described by Gaussians.

\subsection{Comparison of N04 and F05 kinematic analysis}
\label{s:kine}

F05 also published a VizieR catalogue \citep{famaey2004} accompanying their
 paper, based on the CORAVEL radial velocities of 6,691 northern hemisphere
 stars.   Unfortunately, unlike the N04 stars, the F05 stars do not have
 individual stellar ages so F05 cannot be used to re-visit the age-velocity
 dispersion relation of the smooth background and investigate the effects of
 each dynamical stream on the relation.  Instead in this section, we describe
 how F05 decomposed their sample into its constituent kinematic components and
 compare their phase-space substructure with N04 in order to physically
 motivate our assessment of Figs. \ref{fig:uvw25age} and \ref{fig:uvw25hist}
 (and \ref{fig:appendixuvw25age} and \ref{fig:appendixuvw25hist}). 

The basic technique to calculate space velocities is to invert the parallax to estimate the distance.  This is then used to compute the tangential space motion components from the proper motions, which in combination with radial velocities can be transformed to the Galactocentric frame.  Due to the non-linearity of the parallax-distance transformation, if $\sigma_{\pi}/\pi$ is high, the inverse parallax is a biased estimator of the distance and an individual error on a space velocity cannot be derived from a simple first-order linear propagation of $\sigma_{\pi}$.  As all the stars in N04 are relatively nearby, most have parallaxes with $\sigma_{\pi}/\pi$ better than 10 per cent and nearly all are better than 20 per cent.  The availability of photometric distances with an uncertainty of only 13 per cent means the parallax inversion derived distances adopted for N04 stars when $\sigma_{\pi}/\pi <$ 13 per cent only have a very small distance bias \citep{brown1997}, which does not affect the calculation of space velocity errors.

The resulting space velocity diagrams, restricted to the 8,589 single N04
stars, are plotted in the left columns of Figs. \ref{fig:uv}, \ref{fig:uw} and
\ref{fig:vw}
(c.f. N04's fig. 20 of the whole sample).  While the $U - W$
(Fig. \ref{fig:uw}) and $V - W$ (Fig. \ref{fig:vw}) diagrams show a smooth distribution, the $U - V$ diagram (Fig. \ref{fig:uv}) shows abundant phase-space substructure.  It consists of four tilted branches aligned along approximately constant $V$ velocities that resemble classic moving groups or stellar streams named, from top to bottom, the Sirius-UMa, Coma Berenices (or local), Hyades-Pleiades and $\zeta$ Herculis branches.  These branches are only loosely defined as overdensities in phase-space.  N04's straight-forward space velocity derivation does not include assignment of individual stars to these branches, only its position in phase-space.  

  F05 binaries for which no centre-of-mass radial velocity could be estimated were removed from their sample using their binarity flag ($B$).  A star was excluded in their kinematic analysis if it was a spectroscopic binary ($B$ = 0) or a visual binary with no orbit available ($B$ = 5) or an uncertain case, either a spectroscopic binary or supergiant ($B$ = 3) or for a visual binary ($B$ = 8), leaving 6,030 stars (5,311 K and 719 M giants).   Because the F05 giants are at distances approximately an order of magnitude greater than the N04 dwarfs, the F05 distribution of $\sigma_{\pi}/\pi$ is correspondingly less accurate than that of N04.  There are only 786 F05 stars with $\sigma_{\pi}/\pi <$ 10 per cent, which is too small a sample to characterize the K and M giants in the solar neighbourhood.  $\sigma_{\pi}/\pi <$ 20 per cent increases the sample size to 2,774 stars (2,524 K and 250 M giants).  With this accuracy, the distance bias is still very small and the first-order approximation for the calculation of space velocity errors is still valid.  The resulting density of stars in the $U - V$ diagram with $\sigma_{\pi}/\pi <$ 20 per cent reveals overdensities in the same region of phase-space as the N04 branches (see F05's fig. 7).  We therefore consider that the F05 kinematic phase-space structure seen at distances of a few hundred pc is consistent with that seen locally by N04.

To make full use of all their 6,030 stars and to assign individual stars to each phase-space branch in order to investigate their kinematics, F05 applied the Luri-Mennessier method to their data.  This  maximum-likelihood method is based on a Bayesian
approach and is described in detail in \citet{luri1996}.  It requires a model
describing the basic morphological characteristics of the sample (spatial
distribution, kinematics and $M_{V}$) and a model of the selection criteria
used to define the sample.  A total distribution function is built using these
models to describe the observational characteristics of the sample.   This
consists of a linear combination of partial distribution functions, each of
which describes a group of stars called a `base group'.  Each of these
combines a velocity ellipsoid with a Gaussian magnitude distribution and
exponential height distribution perpendicular to the Galactic plane.  The advantage of this phenomenological model is that it can identify and quantify different subgroups in the data, which cannot be easily parametrized using other methods.   A maximum-likelihood fit of the model parameters to the sample yield the model parameters that best represent the sample given the a priori models assumed.  

The expected space velocities deduced by the Luri-Mennessier method of the 6,030 F05 stars are plotted in the right columns of Figs. \ref{fig:uv}, \ref{fig:uw} and
\ref{fig:vw} (c.f. F05's fig. 9).  Modelling this sample with a single base group returns a low maximum likelihood, suggesting to F05 that a single Schwarzschild ellipsoid does not fit the kinematic properties of giant stars in the solar neighbourhood.  The first acceptable solution they found required three base groups: bright giants/supergiants with `young' kinematics, high-velocity stars and `normal' stars.  This last group contained further small-scale structure, which F05 successfully modelled by three more statistically significant base groups.  Although adding more groups produced statistically better solutions, F05 found that these solutions were not stable because they depended too much on the observed values of some individual stars.  F05 identified the resulting six base groups with known kinematic features in the solar neighbourhood (colour-coded in the bottom right panels of Figs. \ref{fig:uv}, \ref{fig:uw} and
\ref{fig:vw}), which are also seen in the N04 phase-space.

Despite all the differences described above, Figs. \ref{fig:uv}, \ref{fig:uw} and \ref{fig:vw} show the N04 and F05 phase-space is remarkably similar, suggesting the F05 kinematic base group assignment describes structures covering several hundred pc in the solar neighbourhood at least to first order, again suggesting the structures are not remnants of single small star clusters.  We therefore consider the N04 branches to have the same physical origin as postulated by F05 for their corresponding base groups.  

\subsection{Comparison of N04 and F05 phase-space}
\label{s:phase}

\begin{figure}
\centering
        \psfig{figure=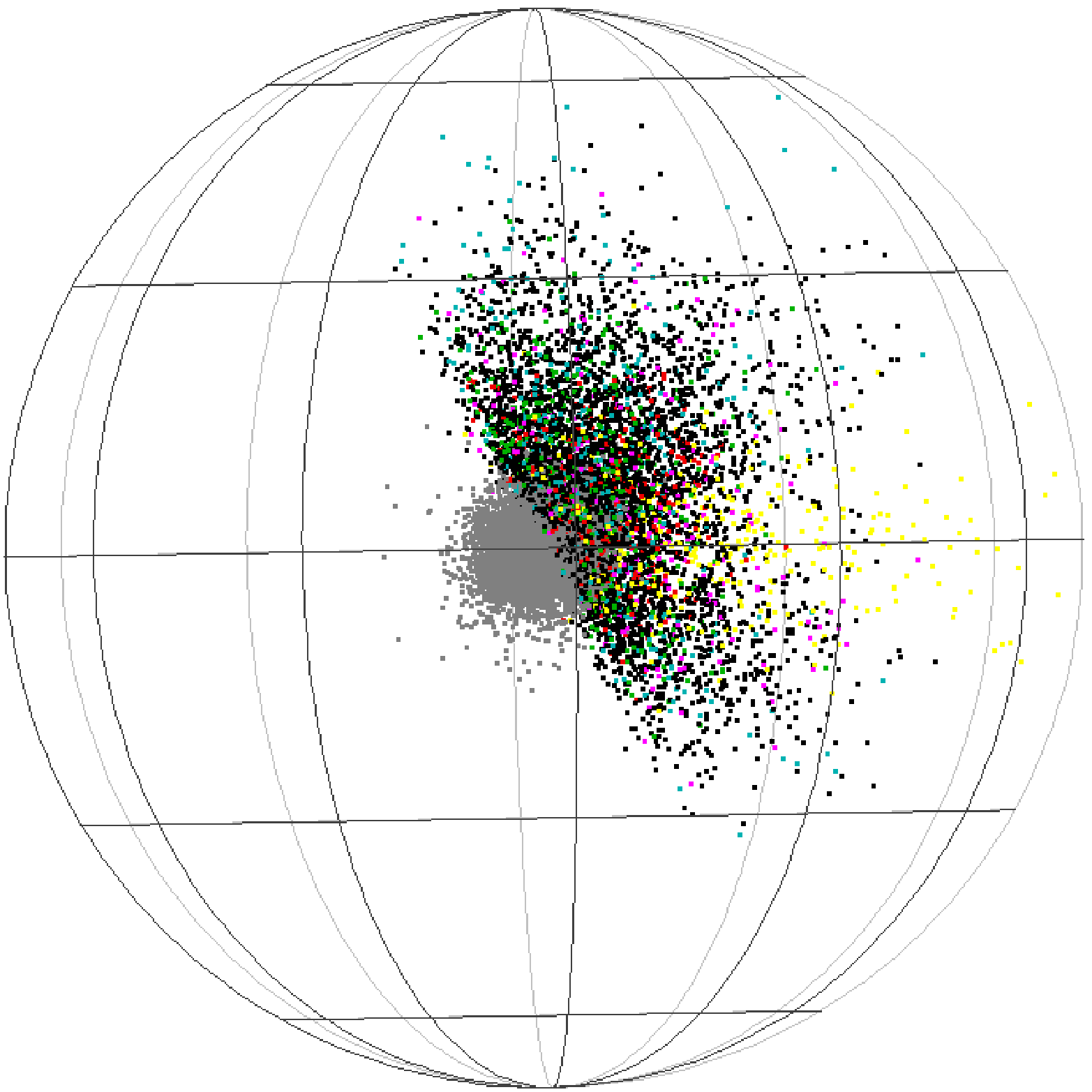,width=1.0\columnwidth}
 \caption{Three dimensional Galactic sky distribution of the 8,589 N04 single
        F-G dwarfs with space velocities (grey) and the 6,030 F05 K-M giants
        with space velocities, including spectroscopic binaries with
        centre-of-mass radial velocities, colour-coded according to their maximum
        likelihood base group
        assignment to known kinematic features of the solar neighbourhood: smooth
  background (black), Sirius moving group (magenta), young
  kinematics (yellow), Hyades-Pleiades supercluster (red),
  Hercules stream (green) and high-velocity stars (blue).
  The spherical polar axes have a radius of 900 pc, centred on the
  Sun.  The viewer is in the Galactic plane, inside the
        solar circle at $l=30^{\circ}$ (the Galactic centre is over the
        viewer's left shoulder).  The viewer's angle has been chosen to
        illustrate the $63^{\circ}$ between the Earth's equator and the
        Galactic plane (longest horizontal line), emphasizing that the
        F05 sample only contains stars visible from the Earth's northern
        hemisphere.  The nearest line of constant Galactic
  longitude is $l=30^{\circ}$ (with the majority
  of the stars behind it).  $l$ increases to the right (anticlockwise) with
        lines of constant Galactic
  longitude every $30^{\circ}$.  All the lines of constant Galactic
  longitude converge at the North Galactic Pole ($b=+90^{\circ}$, top) and at
  the South Galactic Pole ($b=-90^{\circ}$, bottom), with lines of constant Galactic
  latitude every $30^{\circ}$.} 
                \label{fig:3d}
\end{figure}

We compare the space velocity distributions of the N04 and F05 stars in Fig. \ref{fig:uvwgcf}.  This highlights the departures from Gaussianity of the $U$ and $V$ distribution functions caused by the presence of streams compared to $W$, which approximates a Gaussian, due to the tilt of these streams in the $UV$-plane in Fig. \ref{fig:uv}, measured to be $V$ = constant $-$ 0.47$U$ by \citet{skuljan1999}.  A direct comparison between the N04 and F05 distributions is interesting because
Figs. \ref{fig:uv}, \ref{fig:uw} and \ref{fig:vw} show that the N04 substructures are more prominent than the F05 substructures compared to their smooth backgrounds apparently due to the difference in scaleheight between the two samples, illustrated in Fig. \ref{fig:3d}.   The much lower scaleheight of the N04 sample and its more prominent substructure means dynamical streams are more prominent in stars close to the plane due to a mix of dynamics and sample age differences.

Another difference is that there are similar numbers of stars in each of the
F05 streams but the N04 $V$ distribution in Fig. \ref{fig:uvwgcf} shows that
this is not the case in the N04 sample as it is numerically dominated by the
much narrower peak of the Hyades-Pleiades stream at $V \sim -20$ km s$^{-1}$
than in the F05 $V$ distribution.  This is because the N04 completeness out to
$\sim$ 40 pc includes the Hyades cluster itself (visible in Figs. \ref{fig:uv},
\ref{fig:uw} and \ref{fig:vw}).  \citet{famaey2007} showed that in addition to
bound Hyades cluster stars, the Hyades-Pleiades stream is mainly composed of
field-like stars but also partly of coeval stars evaporated from the Hyades
cluster. F05 actually propose that because the cluster is young (600 Myr), it
should not have crossed many spiral structures, suggesting that the same
transient spiral wave that formed it (by boosting star formation in its
primordial gas cloud) could at the same time have given it and nearby field
stars their peculiar velocity in the $UV$-plane.  For any evaporated cluster
stars to get into the much higher scaleheight of the F05 sample requires them
to have experienced phase-mixing and vertical disc heating.  They are more likely to have had their initial birth velocities erased and so are less likely to be identifiable as evaporated cluster stars in the F05 sample.

F05 showed that their dynamical streams span several Gyr in age, which is
supported theoretically by the simulations of \citet{desimone2004}.
Therefore, we expect the F05 streams in Fig. \ref{fig:uvwgcf} to be present in
the N04 sample over many age bins as narrow peaks in the $V$ histograms in
Figs. \ref{fig:uvw25hist} and \ref{fig:appendixuvw25hist}, which allows
us to trace the changing relative strengths of these streams with stellar age.
This suggests that the majority of the peaks in the $V$ histograms in
Figs. \ref{fig:uvw25hist} and \ref{fig:appendixuvw25hist} are in fact dynamical streams rather than noise.  We can also use Fig. \ref{fig:uvwgcf} to interpret the N04 $U$ and $W$ distributions.  It suggests that the many of the minor peaks in these distributions are noise.  However, like in $V$, the F05 streams do not share similar mean $U$ velocities so the N04 $U$ distribution is also complex and non-Gaussian.   We conclude that N04's treatment of the $U$ and $V$ distributions with a straight-forward dispersion was an inadequate parametrization of these complex distribution functions.  The interpretation of an age-velocity dispersion relation for $U$ and $V$ is non-trivial and beyond the scope of this paper.  

Nevertheless, these complex distribution functions raise the interesting issue as to what constitutes disc heating.  In the introduction, we defined it as processes that change stellar velocities in any direction such that the velocity dispersion increases.  If this heating is random in direction it will fill and shape the velocity ellipsoid slowly over time.  However, if processes like those responsible for dynamical streams (transient spiral waves and/or resonances) overpopulate a specific subregion of phase-space, then the concentration of these stars relative to the background can actually change the overall components of the velocity dispersion by weighting them to the mean velocities of the streams.  This scenario is illustrated in Table 1.  It reveals that the F05 background dispersions are systematically slightly larger than the dispersions of the background and thin disc streams combined.  This shows that the F05 streams decrease the dispersions of the background and thin disc streams combined by weighting them to the mean velocities of the streams so that the combined dispersion differs slightly from the background dispersion.  Therefore, the mechanism responsible for the F05 thin disc streams could be considered a disc `cooling' agent.  The inclusion of the Hercules stream and high-velocity stars increases the overall dispersion to greater than the background alone so they could be considered as `heating' agents.  However, this classification is specific to the F05 sample.  The dominance of the N04 streams over the N04 background complicates the $U$ and $V$ distributions more than in F05.  The differing strengths with stellar age of the streams, makes it difficult to estimate whether the overall $\sigma_{U}$ and $\sigma_{V}$ will be heated or cooled by the presence of streams compared to the background as a function of time, without prior knowledge of the properties of the N04 background, which are non-trivial to derive given the dominance of the streams.  Therefore, we are unable to generalize which specific mechanisms constitute disc heating or cooling because it depends on the sample considered.

\begin{table}
\label{tab:streams}
  \caption{Comparison of N04 and F05 velocity dispersions (in km s$^{-1}$) of combinations of different F05 kinematic groups: smooth background (B), Sirius moving group (S), young
  kinematics (Y), Hyades-Pleiades supercluster (HyPl),
  Hercules stream (He) and high-velocity stars.  (Each dispersion is an 
outlier-resistant measure of the dispersion about the centre of the
velocity distribution calculated by using the
median absolute deviation as the initial estimate, then each star is weighted using Tukey's Biweight. For a Gaussian
distribution, this is identical to the standard deviation.)}
  \centering
  \begin{tabular}{@{}lccc@{}} 
   \hline 
Dispersion & $\sigma_{U}$ & $\sigma_{V}$ & $\sigma_{W}$ \\
\hline 
N04 all & 32.6 $\pm$ 0.2 & 20.0 $\pm$ 0.2 & 15.1 $\pm$ 0.1\\
\hline 
B & 33.8 $\pm$ 0.4 & 17.4 $\pm$ 0.2 & 17.8 $\pm$ 0.2\\ 
Si, Y, HyPl & 22.3 $\pm$ 0.5 & 12.2 $\pm$ 0.3 & ~7.0 $\pm$ 0.2\\
B, Si, Y, HyPl & 31.8 $\pm$ 0.3 & 16.4 $\pm$ 0.2 & 15.7 $\pm$ 0.2\\
B, Si, Y, HyPl, He & 33.7 $\pm$ 0.3 & 19.6 $\pm$ 0.2 & 15.8 $\pm$ 0.1\\
F05 all & 35.6 $\pm$ 0.3 & 22.0 $\pm$ 0.2 & 17.1 $\pm$ 0.2\\
\hline 
\end{tabular}
\end{table}

N04 excluded the youngest age bin from their power law fit to avoid biases due
to unrelaxed young structures.  The non-Gaussianity of some of the $W$
histograms of the young thin disc (0.5 $<$ age $\le$ 2.8 Gyr) bins (see
Fig. \ref{fig:appendixuvw25hist}) suggests that
the N04 sample does not vertically relax until the stars are older than this age range.  The $W$
histograms of the 2.8 $<$ age $\le$ 4.6 Gyr (see Fig. \ref{fig:uvw25hist}) and 4.6
$<$ age $\le$ 7.6 Gyr bins (see
Fig. \ref{fig:appendixuvw25hist}), representing the old thin disc, are more Gaussian and thus can be considered relaxed.  We conclude that N04's treatment of the $W$ distribution with a straight-forward dispersion may not be an adequate parametrization of the young thin disc but it is adequate for the old thin disc.  This does not affect their general result because $\sigma_{W}$ must increase with time in order for the older thin disc to have a higher scaleheight than the young thin disc.  It does not affect our aims here because we are only concerned with the functional form of the relation for the old thin disc.

It has long been known that traditional moving groups are indistinguishable in the vertical direction because the $W$ velocity distribution function appears to be well phase-mixed \citep{dehnen1998b}.  This is theoretically expected as phase-mixing of vertical space motions is more efficient than for the horizontal space motions because, for a realistic disc profile, the vertical frequency is a strong function of vertical energy and the vertical dynamical time is a factor of two shorter than the horizontal orbital period.  Consequently, $W$ motion is decoupled from $U$ and $V$ motions, suggesting that dynamical streams should not affect a star's $W$ velocity.  Table 2 in F05 suggest that all the streams share similar mean $W$ ($\langle W \rangle$).  We test this in the following section.

\subsection{Stellar warp of the Galactic disc}
\label{s:warp}

\begin{figure*}
\begin{minipage}{170mm}
\centering
        \psfig{figure=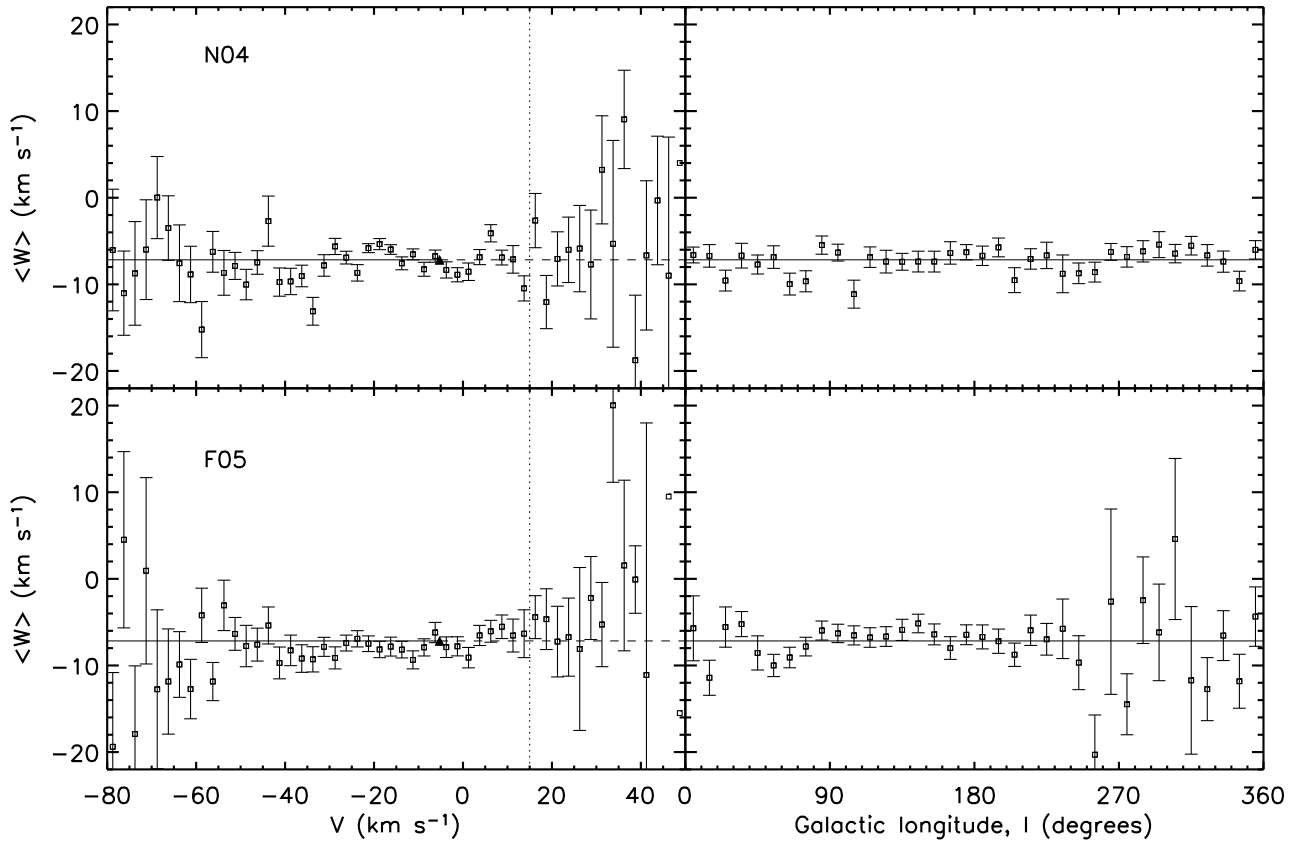,width=1\columnwidth}
 \caption{Robust mean vertical motion ($\langle W \rangle$) as a function of $V$
        (left column, same $V$ plot range, velocity bin centres and 2.5 km s$^{-1}$ velocity bin widths as fig. 6 in 
        \citealt{dehnen1998b}) and as a function of Galactic longitude, $l$ (right column) for
        the N04 single F-G dwarfs
         and the F05
        K-M giants.  8,386 N04 stars are included in
        the top left plot out of a possible 8,589 (98 per cent), whereas all are included in the top right plot. 5,852 F05 stars are included in
        the bottom left plot out of a possible 6,030 (97 per cent), whereas all are included in the bottom right plot. The triangle
        indicates the velocity of the Local Standard of Rest (LSR) with respect to
        the Sun: ($V_{\mathrm{LSR}}$,$W_{\mathrm{LSR}}$) =
        ($-V_{\odot}$,$-W_{\odot}$) = (-5.25,-7.17) km s$^{-1}$ \citep{dehnen1998a}.  The
        horizontal solid
        lines along $W_{\mathrm{LSR}}$ out to the LSR represents a flat Galactic
        disc out to the solar circle.  The horizontal dashed lines outwards from the LSR
        represent a model of the flat disc continuing outside the
        solar circle instead of a stellar warp in the disc, as proposed by
        \citet{dehnen1998b}.  The $\chi^{2}$ tests compare the data to the
        model to the right of the vertical dotted lines at $V$ = 15 km
        s$^{-1}$ (set to exclude the maximum $V$ bins containing Sirius dynamical
        stream stars).  Data points without error bars on $\langle W \rangle$ only
        contain one star and are thus excluded from the $\chi^{2}$ tests (as
        is the empty F05 bin $42.5 < V \le 45.0$ km
        s$^{-1}$).}  
                \label{fig:warp}
                \end{minipage}
\end{figure*}

\citet{dehnen1998b} found that his outermost contours of phase-space density are
skewed relative to the $W$ axis in the sense that more stars moving faster than the
Local Standard of Rest (LSR, $+V$) are also moving upward
with respect to the LSR ($+W$) than downwards ($-W$).  His fig. 6 shows that
$\langle W \rangle$ is approximately constant at the
$W_{\mathrm{LSR}}$ value for 
$V \lesssim$ 10 km s$^{-1}$ but curves upward for $V \gtrsim$ 10 km s$^{-1}$.  He interprets this curve as the signature of a
stellar warp in the Galactic disc, which starts at the solar circle.   If there are lots of N04 stars in the warp and these stars are included in our
analysis, the departure from a constant
$\langle W \rangle$ would artificially increase $\sigma_{W}$.  This would bias
the true $\sigma_{W}$ so that it could appear as though there is more disc
heating than in the assumed flat disc case (approximately constant
$W_{\mathrm{LSR}}$ with respect to $V$).  

To check whether this source of $\sigma_{W}$ bias is present in the N04 data
set, we reproduce fig. 6 from \citet{dehnen1998b} for the N04 and F05 samples in the left column of
Fig. \ref{fig:warp}.   Contrary to his claim, we do not find a signature of a
stellar warp in the Galactic disc in either data set.
Fig. \ref{fig:uv} shows that the number of stars decreases with $+V$ in both
data sets, which is reflected in the large error bars in Fig. \ref{fig:warp}
at high values of $+V$. 

 The N04 sample in the top left panel of Fig. \ref{fig:warp} is averaged over
 the entire sky.  HI observations show that the Galactic disc is flat out to
 approximately the solar circle and then the warp turns up towards the North
 Galactic Pole in the $l \sim 90^{\circ}$ direction, while turning towards the
 South Galactic Pole in the $l \sim 270^{\circ}$ direction.  If the N04 and
 F05 stars share the gas kinematics of the warp, then their participation in
 the warp would be seen by $\langle W \rangle > 0$ towards $l \sim 90^{\circ}$
 and $\langle W \rangle < 0$ towards $l \sim 270^{\circ}$.  It is conceivable
 that if the warp is symmetric in amplitude about the plane and symmetric in
 $l$, then its signature could cancel itself out and result in a flat $\langle
 W \rangle$ as a function of $V$ in the top left panel of Fig. \ref{fig:warp}.  Plotting $\langle W \rangle$ as a function of $l$ in the right column of Fig. \ref{fig:warp} confirms this is not the case and that there is no sinusoidal signature of the warp in $l$.  The F05 sample becomes noisy at $l > 230^{\circ}$ because of incomplete coverage of the Galactic sky (see Fig. \ref{fig:3d} and F05 fig. 6).
 
 \begin{table}
  \centering
\label{tab:warp}
  \caption{$\chi^{2}$ goodness-of-fit test results between the data
    and model in Fig. \ref{fig:warp}.  $n$ is the number of data points included in the model fit.  $n_{c}$ is the
number of constraints on each model fit (only one to normalize each model to each
data set).   The number of degrees of freedom is $\nu = n - n_{c}$.  The $\chi^{2}$ probability distribution function for $\nu$ degrees of freedom,
$Q(\chi^{2}|\nu)$, is the standard statistical significance of the $\chi^{2}$
test ($Q(\chi^{2}|\nu) = 1 - P(\chi^{2}|\nu)$).} 
  \begin{tabular}{@{}lcccccc@{}} 
  \hline
Sample  & $n$ & $n_{c}$ & $\nu$ & $\chi^{2}$ & $\chi^{2}/\nu$ & $Q(\chi^{2}|\nu)$\\
\hline 
N04 $V>15$ & 13 & 1 & 12 & 19.0 & 1.58 & 0.089\\
F05 $V>15$ & 11 & 1 & 10 & 16.5 & 1.65 & 0.087\\
\hline 
N04 all $l$ & 36 & 1 & 35 & 45.2 & 1.29 & 0.117\\
F05 all $l$ & 36 & 1 & 35 & 47.1 & 1.35 & 0.083\\
\hline
\end{tabular}
\end{table}

Table 2 shows that $Q(\chi^{2}|\nu) > 0.003$ ($P(\chi^{2}|\nu) < 0.997$) for all the tests so both data sets accept
 the flat disc model at the $3\sigma$ statistical significance level. 
 However, the model is not strongly accepted as would be indicated by high
$Q(\chi^{2}|\nu)$ values.  This is because for each data set, there is one $V$ bin at similar $V$ values ($32.5 < V \le
        35.0$ and $35.0 < V \le 37.5$ km
        s$^{-1}$ for F05 and N04 respectively) for which the $\langle W \rangle$ is
more than 2$\Delta\langle W \rangle$ away from the flat disc model of constant
$W_{\mathrm{LSR}}$ (i.e. no warp).  It is these bins that cause the low
$Q(\chi^{2}|\nu)$ values in Table 2.  Respectively, the 6 and 4
stars in these bins do not share similar $W$ velocities as would be expected
if they were part of a
warp or stream.  These bins have very wide
ranges of $W$ velocities, perhaps because one or more of these stars are
members of the thick disc or halo.  The small number of stars in each
bin results in the data points being offset from the rest.

Therefore, contrary to \citet{dehnen1998b}, we do not find a signature of the
        stellar warp in the Galactic disc in the N04 or F05 samples.  This
        means that the N04 sample has an approximately constant $\langle W
        \rangle$.  The small ($\sim$ 2 km
        s$^{-1}$) excursions from $W_{\mathrm{LSR}}$ as a function of $V$ do coincide with the $V$ values of the streams in the N04 sample.  These small net vertical motions are presumably due to the young unrelaxed stars evaporating from star clusters rather than field stars caught in the dynamical streams.

The lower scaleheight of the N04 sample in Fig. \ref{fig:3d} means its smooth
background $\sigma_{W}$ will be less than the F05 background $\sigma_{W}$ in
Table 2.  \citet{famaey2007} showed that $\sim$ 15 per cent of the N04
Hyades-Pleiades stream is evaporated coeval Hyades cluster stars.  Assuming an
initial velocity dispersion $\sigma_{0} \sim$ 10 km s$^{-1}$, the
evaporated Hyades stars will manifest themselves as a clump in the velocity
distribution function that is narrow in $V$ (because all the stars have
identical angular momentum) but has size $\gtrsim \sigma_{0}$ in $U$ and $W$ \citep{woolley1961}.
Therefore, the N04 Hyades-Pleiades stream should have a larger $\sigma_{W}$ than
the purely dynamical F05 Hyades-Pleiades stream (this may also be the case for
the other streams).  Hence, the N04 background $\sigma_{W}$ is likely to be
more similar to the N04 thin disc streams $\sigma_{W}$ than in the F05 case.
Thus the overall N04 $W$ distribution is a sum of similar $\sigma_{W}$ and
similar $\langle W \rangle$ Gaussians, which is a Gaussian with similar
$\sigma_{W}$ and similar $\langle W \rangle$.  Having established that the only component of the age-velocity dispersion relation that can be simply measured locally is the vertical component, the rest of the paper concentrates on re-visiting N04's age-$\sigma_{W}$ relation.

\subsection{Reproducing the N04 age-$\sigma_{W}$ relation}
\label{s:age25}

\begin{figure*}
\begin{minipage}{170mm}
\centering
        \psfig{figure=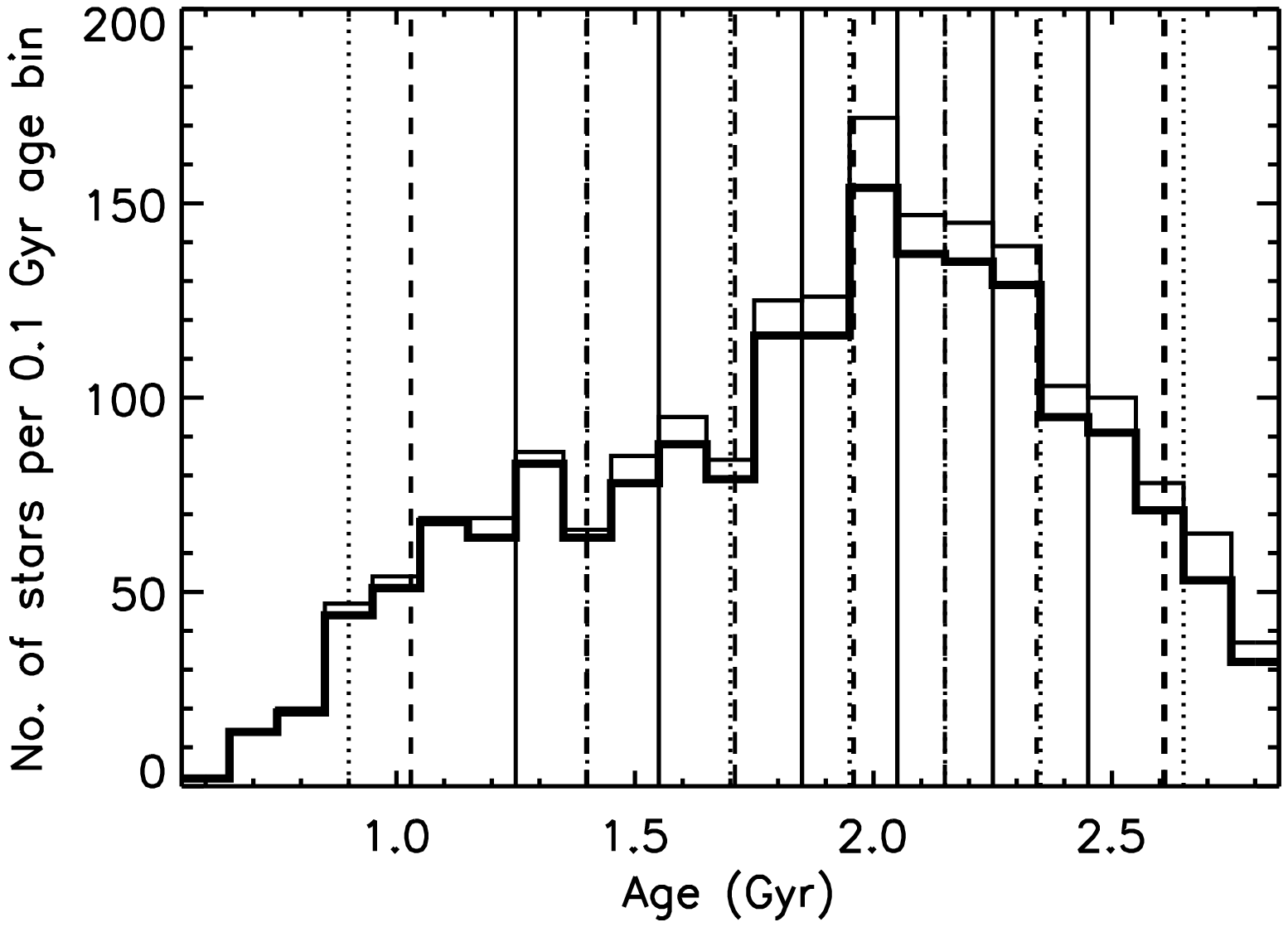,width=0.45\columnwidth}
        \psfig{figure=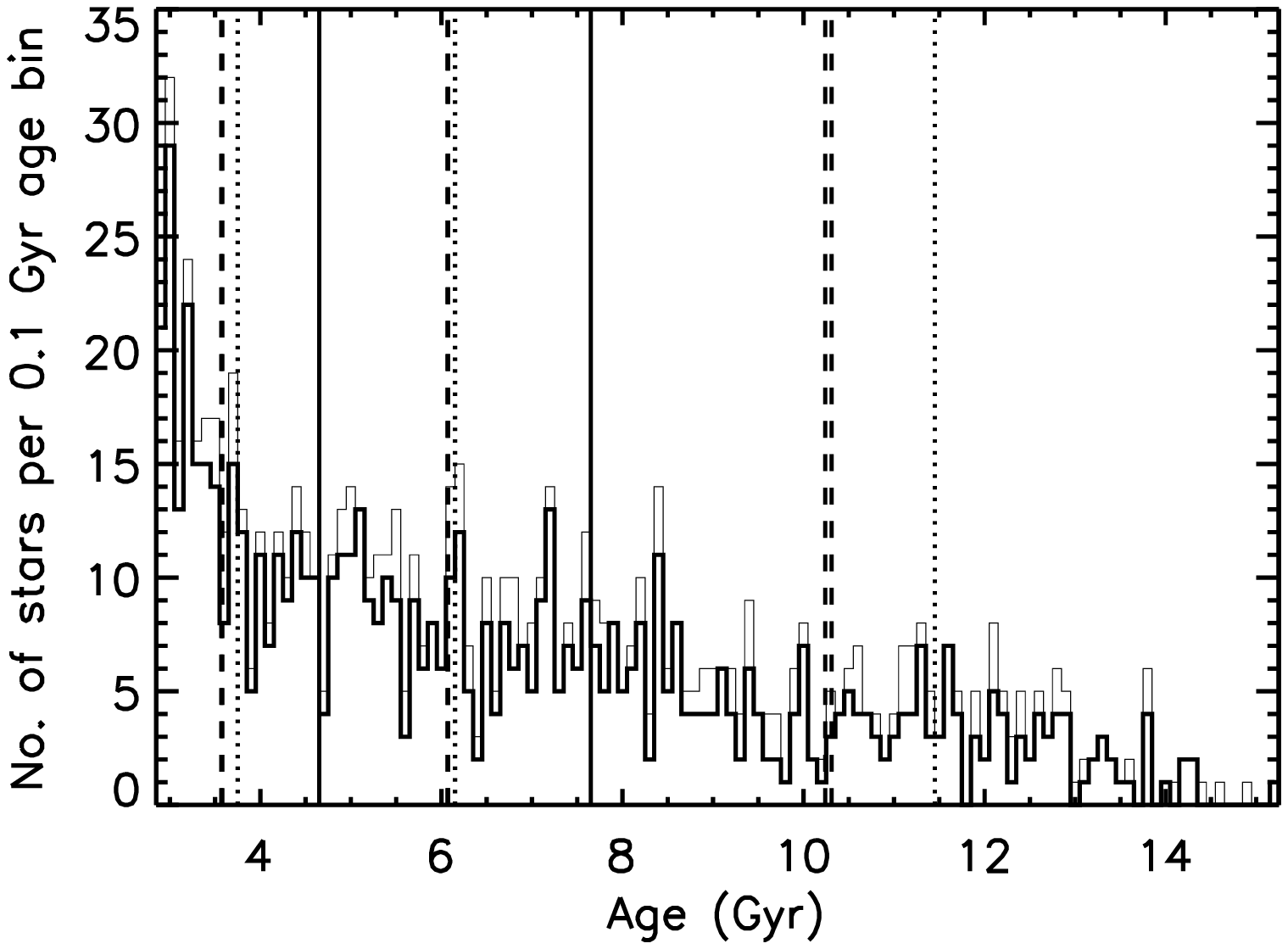,width=0.45\columnwidth} 
\caption{Discrete age distribution of the N04 single stars with relative age errors $<$
        25 per cent, split between the youngest 7 age bins from Fig. \ref{fig:appendixuvw25hist}
        (left) and the oldest 3 age bins from Fig. \ref{fig:appendixuvw25hist}
        (right).  The thin lines correspond to the whole sample and the thick
        lines are the subsample excluding the Hercules stream.  Because the N04 catalogue stellar ages are provided to the nearest 0.1
        Gyr, each 0.1 Gyr histogram bin only contains stars with the exact age
        that its histogram step is centred on e.g. the youngest histogram bin,
        centred on 0.6 Gyr, only contains stars with exactly 0.6 Gyr ages but the histogram step spans 0.55 to 0.65
        Gyr.  The vertical solid lines (including the vertical edges of the
        plots) are the boundaries to the age bins from
        Fig. \ref{fig:appendixuvw25hist}.  The dotted lines are the 
        central ages of each age bin.  The robust mean ages
        within each age bin for the whole sample (thin dashed lines) and subsample (thick dashed
        lines) are indisguishable apart from the oldest age bin.}
                \label{fig:age25}
                \end{minipage}
\end{figure*}

Fig. \ref{fig:uvwgcf} shows that the Hercules stream has a larger $\sigma_{W}$ than the other F05 streams.  This may be due to its different dynamical origin from the other streams: perhaps the Galactic bar.  \citet{bensby2007} showed that the Hercules stream includes both younger and old disc stars and they suggest some thick disc stars may be included.  The Hercules stream $\sigma_{W}$ may be larger due to the thick disc stars in the stream retaining their characteristically higher $\sigma_{W}$, which is a property of the thick disc.  Alternatively, the higher $\sigma_{W}$ may be a dynamical signature of the bar, imparted to both its constituent thin and thick disc stars as they interacted with the bar.  If it is the latter and the N04 Hercules stream $\sigma_{W}$ is sufficiently different from the N04 smooth background and thin disc streams, then the thin disc Hercules stream stars could bias the age-$\sigma_{W}$ relation.   The smooth $W$ distributions in Figs. \ref{fig:uw} and \ref{fig:vw} suggest $\sigma_{W}$ is a valid
description of the $W$ distribution function, unbiased by kinematic
substructure because the $W$ distribution is well phase-mixed. By deriving the age-$\sigma_{W}$ relation both with and without the Hercules stream, we can simultaneously test whether the relation is biased by the stream and whether the $W$ distribution is well phase-mixed, implying the presence of the thin disc streams does not affect the relation.

We use the most extreme F05 Hercules stream stars in $U$ and $V$ to define the
N04 Hercules stream $UV$ region in Fig. \ref{fig:uv}.  Figs. \ref{fig:uw} and
\ref{fig:vw} are also used to identify high-velocity stars within the $UV$ region of the Hercules stream that have
$W$ velocities that are outside the Hercules stream $W$ region and
thus are more likely to be part of the thick disc velocity ellipsoid.  This technique will
only identify high-velocity stars with more extreme thick disc or halo
kinematics.  Therefore, the removal of Hercules stream stars will also
remove some high-velocity stars but this is of no consequence to our
study.    These stars are colour-coded in Figs. \ref{fig:uv}, \ref{fig:uw} and \ref{fig:vw}.  Every figure from here onwards both includes and excludes Hercules stream stars.

\begin{figure*}
\centering
\begin{minipage}{170mm}
\includegraphics[width=1.0\columnwidth]{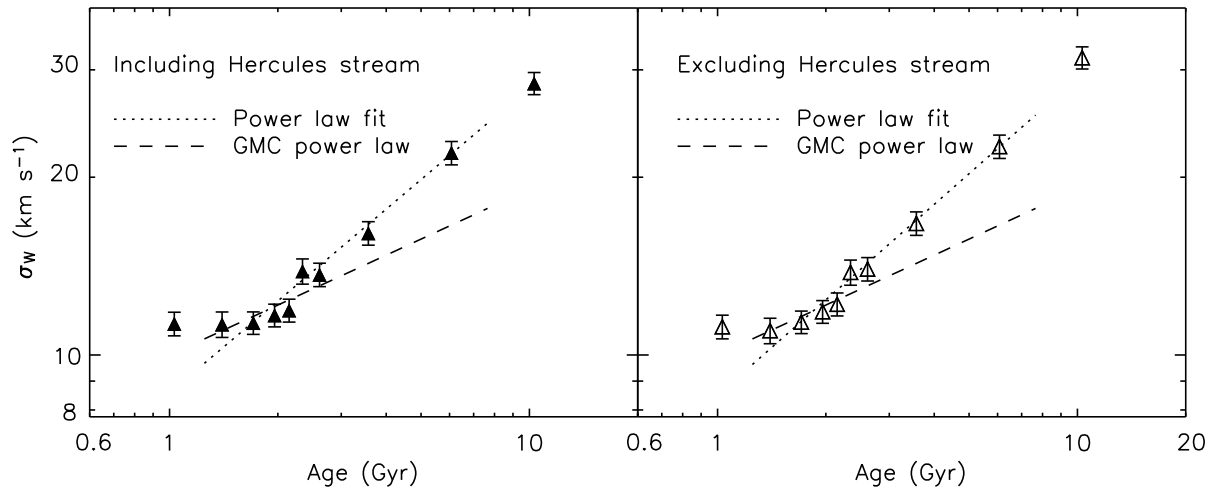}
\caption{N04 logarithmic age-$\sigma_{W}$ relation: robust $\sigma_{W}$ of N04 single stars with relative age errors $<$
        25 per cent in each of the 10 age bins in Fig. \ref{fig:appendixuvw25hist}
as a function of the robust mean age of each
        bin including (left, filled triangles) and excluding (right, open triangles) the Hercules
        stream. As in N04, the youngest and oldest age bins have been
         excluded from the power law fit to avoid biases due to unrelaxed young
         structures and thick disc stars respectively.}
                \label{fig:log25}
\end{minipage}
\end{figure*}

Fig. \ref{fig:age25} shows that a robust (bisquare weighted) mean age of an
age bin is sometimes quite different to the central age of that bin.  By comparison with N04's fig. 31, we find we are best able to reproduce their age-$\sigma_{W}$ relation using both robust ages and robust dispersions in Fig. \ref{fig:log25}.   The error bars in Fig. \ref{fig:log25} are the errors in calculating a standard deviation: 

\begin{equation}
 \Delta\sigma_{W}=\frac{\sigma_{W}}{\sqrt{2N}},
 \label{equ:error}
 \end{equation}

\noindent where $N$ is the number of stars in each age bin used to calculate
$\sigma_{W}$.  The logarithmic spacing of the age bins in Fig. \ref{fig:log25} is a natural consequence of N04's decision to have equal numbers of stars in each of their age bins because the distribution of stellar ages is logarithmic.  This also naturally leads N04 (and us) to fit a power law:

\begin{equation}
\sigma_{W} = a\,\mathrm{age}^{k},
 \label{equ:powerlinear}
 \end{equation}

\noindent where $a$ is a constant and $k$ is the scaling (heating) exponent.  Taking the log of
both sides of equation \ref{equ:powerlinear} gives

\begin{equation}
\mathrm{log}_{10}(\sigma_{W})=k\,\mathrm{log}_{10}(\mathrm{age}) + \mathrm{log}_{10}(a),
 \label{equ:powerlog}
 \end{equation}

\noindent which produces a linear relationship where $k$ is now the slope and log$_{10}(a)$ is
the intercept.  A linear least-squares fit gives $k = 0.48 \pm 0.26$ (including the Hercules stream, $k$ = 0.50
$\pm$ 0.25 excluding Hercules stream), which
is similar to N04's $k = 0.47 \pm 0.05$.  However, we can only reproduce their quoted uncertainty by excluding errors from the fit: $k = 0.45 \pm 0.04$.   We are thus able to
reproduce and verify the N04 age-$\sigma_{W}$ relation using the rounded up catalogue data.  

These values of $k$ are close to the $k$ = 0.50 predicted by \citet{hanninen2002} for vertical disc heating by halo black holes of mass $1 \times 10^{7}$ M$_{\odot}$.  The goodness of fit of these power laws are therefore a test of how well the data are represented by this source of heating.  \citet{hanninen2002} also predict $k$ = 0.26 for the vertical heating exponent by GMCs, where the constant $a$ in equation \ref{equ:powerlinear} needs to be set by observations.  We have plotted their power law prediction in Fig. \ref{fig:log25}, setting $a = \sigma_{0}$ = 10 km s$^{-1}$ so that at the youngest age included in the fits (1.25 Gyr), $\sigma_{W} \approx$ 10.6 km s$^{-1}$.

\begin{figure*}
\centering
\begin{minipage}{170mm}
\includegraphics[width=1.0\columnwidth]{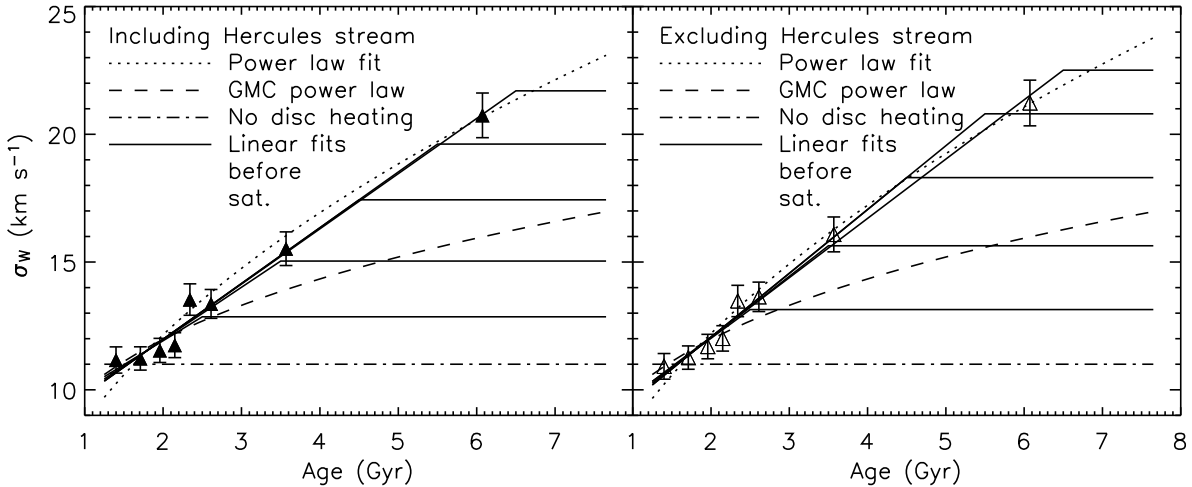}
\caption{N04 linear age-$\sigma_{W}$ relation: robust $\sigma_{W}$ of N04 single stars with relative age errors $<$
        25 per cent in the 8 age bins included in the power law fit in
        Fig. \ref{fig:log25} as a function of the robust mean age of each
        bin including (left, filled triangles) and excluding (right, open triangles) the Hercules
        stream.  The dashed lines are models representing no disc heating,
        where the constant saturation level is set by the dispersion in the
        youngest bin in Fig. \ref{fig:log25}.  The solid lines are linear fits to each set of data points up
        to 2.5, 3.5, 4.5, 5.5 and 6.5 Gyr.  Thereafter these models saturate at
        constant dispersions.}
                \label{fig:linear25}
\end{minipage}
\end{figure*}

The linear age-$\sigma_{W}$ plane in
Fig. \ref{fig:linear25} highlights the unequal spacing in age of the age bins as a result of the N04 procedure requiring
the same number of stars in each bin.  The densely distributed age
bins at the young end confirm the presence of an age-$\sigma_{W}$ relation for
the young thin disc.   However, N04 do not consider whether any other models
fit their data better than a power law.  They assume a power law is the
best-fitting model and interpret this as evidence of continued disc heating throughout its lifetime.  The relative age error
of $<$ 25 per cent translates to sparse sampling of the old thin disc
(two data points) and the thick disc (one data point).  These are the only
three age bins between 2.9 and 15.2 Gyr and only the former two bins are
included in the power law fit.  Therefore, the power law is not as tightly
constrained at the old end as it is at the young end.  

We test N04's assumption that
a power law is the best-fitting model by generating alternative time dependencies of disc heating models
(overlaid in Fig. \ref{fig:linear25}) and comparing them to the data.  Our main
aim is to establish whether the age-$\sigma_{W}$ relation saturates for older
stars.  This means we are not so concerned with the exact functional form of
the relation younger than a hypothetical saturation age.  Thus we simply
define models as linear fits of the data points up to a hypothetical
saturation age, after which $\sigma_{W}$ is constant with time, set by the
$\sigma_{W}$ at the hypothetical saturation age.  The number of constraints on each power law model fit ($n_{c}$) is 2 because the number of degrees of freedom ($\nu$) is reduced by the
two constraints of the fit: the slope and
intercept.  This is also the case for each linear fit before
saturation model but $\nu$ is further reduced by one more constraint: the
choice of saturation epoch.  The $\chi^{2}$ statistic for our purpose of comparing the observed distribution of
$\sigma_{W}$ to a model is 

\begin{equation}
\chi^{2} = \sum_{i=1}^n\left(\frac{\sigma_{W_{i}}^{O} - \sigma_{W_{i}}^{E}}{\Delta\sigma_{W_{i}}^{O}}\right)^{2},
\label{equ:chi2sigmaw}
\end{equation}

\noindent where $\sigma_{W_{i}}^{O}$ is the observed $\sigma_{W}$ of the $i$th age bin,
measured as a robust $\sigma_{W}$, and $\sigma_{W_{i}}^{E}$ is the robust
$\sigma_{W}$ expected according to a model. 

 \begin{table*}
  \centering
  \begin{minipage}{170mm}
 \centering
\label{tab:age25}
  \caption{$\chi^{2}$ goodness-of-fit test results between the N04 age binning
    data
    and disc heating models in Fig. \ref{fig:linear25}.  
$n$ is the number of data points included in the model fit.
$n_{c}$ is the
number of constraints on the model fit. 
 The number of degrees of freedom is $\nu = n - n_{c}$.  The $\chi^{2}$ probability distribution function for $\nu$ degrees of freedom,
$Q(\chi^{2}|\nu)$, is the standard statistical significance of the $\chi^{2}$
test ($Q(\chi^{2}|\nu) = 1 - P(\chi^{2}|\nu)$). If
    $Q(\chi^{2}|\nu) > 0.003$ ($P(\chi^{2}|\nu) < 0.997$), the data accept the model at the
    $3\sigma$ statistical significance level. If $Q(\chi^{2}|\nu) <$ 0.003 ($P(\chi^{2}|\nu) > 0.997$), the data reject the model at the
    $3\sigma$ statistical significance level.} 
  \begin{tabular}{@{}lrrrlrrrl@{}} 
  \hline
Model & $n$ & $n_{c}$  & $\nu$ & Hercules & $\chi^{2}$ & $\chi^{2}/\nu$ & $Q(\chi^{2}|\nu)$ & $3\sigma$ significance\\
\hline
Power law fit & 8 & 2 & 6 & Included & 9.0 & 1.5 & 0.176 & The data accept the model\\
                    & & & & Excluded & 4.8 & 0.8 & 0.571 & The data accept the model\\
\hline
GMC power law  & 8 & 2 & 6 & Included & 41.5 & 6.9 & $10^{-7}$ & The data reject the model\\
                                         & & & & Excluded & 49.6 & 8.3 & $10^{-7}$ & The data reject the model\\
 \hline
No disc heating & 8 & 1 & 7 & Included & 210.1 & 30.0 & $<10^{-16}$ & The data reject the model\\
                              & & & & Excluded & 230.3 & 32.9 & $<10^{-16}$ & The data reject the model\\
\hline
Linear fit before saturation at 2.5 Gyr & 8 & 3 & 5 & Included & 103.1 & 20.6 & $<10^{-16}$ & The data reject the model\\
                                              & & & & Excluded & 103.5 & 20.7 & $<10^{-16}$ & The data reject the model\\
\hline
Linear fit before saturation at 3.5 Gyr  & 8 & 3 & 5 & Included & 47.8 & 9.5 & $10^{-9}$ & The data reject the model\\
                                               & & & & Excluded & 41.9 & 8.4 & $10^{-7}$ & The data reject the model\\
\hline
Linear fit before saturation at 4.5 Gyr & 8 & 3 & 5 & Included & 19.0 & 3.8 & 0.002 & The data reject the model\\
                                              & & & & Excluded & 13.1 & 2.6 & 0.022 & The data accept the model\\
\hline
Linear fit before saturation at 5.5 Gyr & 8 & 3 & 5 & Included & 6.3 & 1.3 & 0.279 & The data accept the model\\
                                              & & & & Excluded & 2.7 & 0.5 & 0.752 & The data accept the model\\
\hline
Linear fit before saturation at 6.5 Gyr & 8 & 3 & 5 & Included & 5.8 & 1.2 & 0.322 & The data accept the model\\
                                              & & & & Excluded & 4.8  & 1.0 & 0.445 & The data accept the model\\
\hline
\end{tabular}
\end{minipage}
\end{table*}

As expected, Figs. \ref{fig:log25} and \ref{fig:linear25} show that the largest effect in removing the Hercules stream from the N04 binning sample is seen in the oldest thick disc age bin, which is not included in the age-$\sigma_{W}$ relation.  Excluding the Hercules stream stars removes Hercules stream thin disc stars from the thick disc region of phase-space and so the 
thick disc $\sigma_{W}$ increases slightly.  Figs. \ref{fig:log25} and \ref{fig:linear25} and Table 3 show that removing the Hercules stream has a negligible effect on the age-$\sigma_{W}$ relation.   This may suggest that there are insufficient numbers of Hercules stars in each age bin for the hypothetically higher Hercules $\sigma_{W}$ to affect the relation.  It also suggests that dynamical streams are well phase-mixed in $W$, supporting our argument that the age-$\sigma_{W}$ relation can be constrained from the N04 data.

Table 3 shows that the N04 age-binned data (including the Hercules stream)
accept the power law and thus halo black holes are not ruled out as vertical
disc heating sources.  As expected the data strongly reject a no disc heating
model and linear fits before saturation at 2.5 and 3.5 Gyr models.  The data
also reject the GMC heating power law.  However, the $Q(\chi^{2}|\nu)$ values
for this model are very sensitive to the initial $\sigma_{W}$.  As mentioned
earlier, if the unrelaxed early age bins actually have larger values of
$\sigma_{W}$, then the \citet{hanninen2002} model could perhaps be accepted by
the data.  We confirm one of N04's key results that disc heating does not
saturate at an early stage.  The linear fit before saturation at 4.5 Gyr model
is only marginally rejected by the data including the Hercules stream, while
the data excluding the Hercules stream accept this model.  Models with linear
fits before saturation at 5.5 and 6.5 Gyr are accepted by the data including
and excluding the Hercules stream.  Hence, the N04 data do not require a power
law and actually statistically prefer saturation after $\sim$ 4.5 Gyr.  

\subsection{Higher resolution age binning}
\label{s:age1}

\begin{figure*}
\begin{minipage}{170mm}
\centering
        \psfig{figure=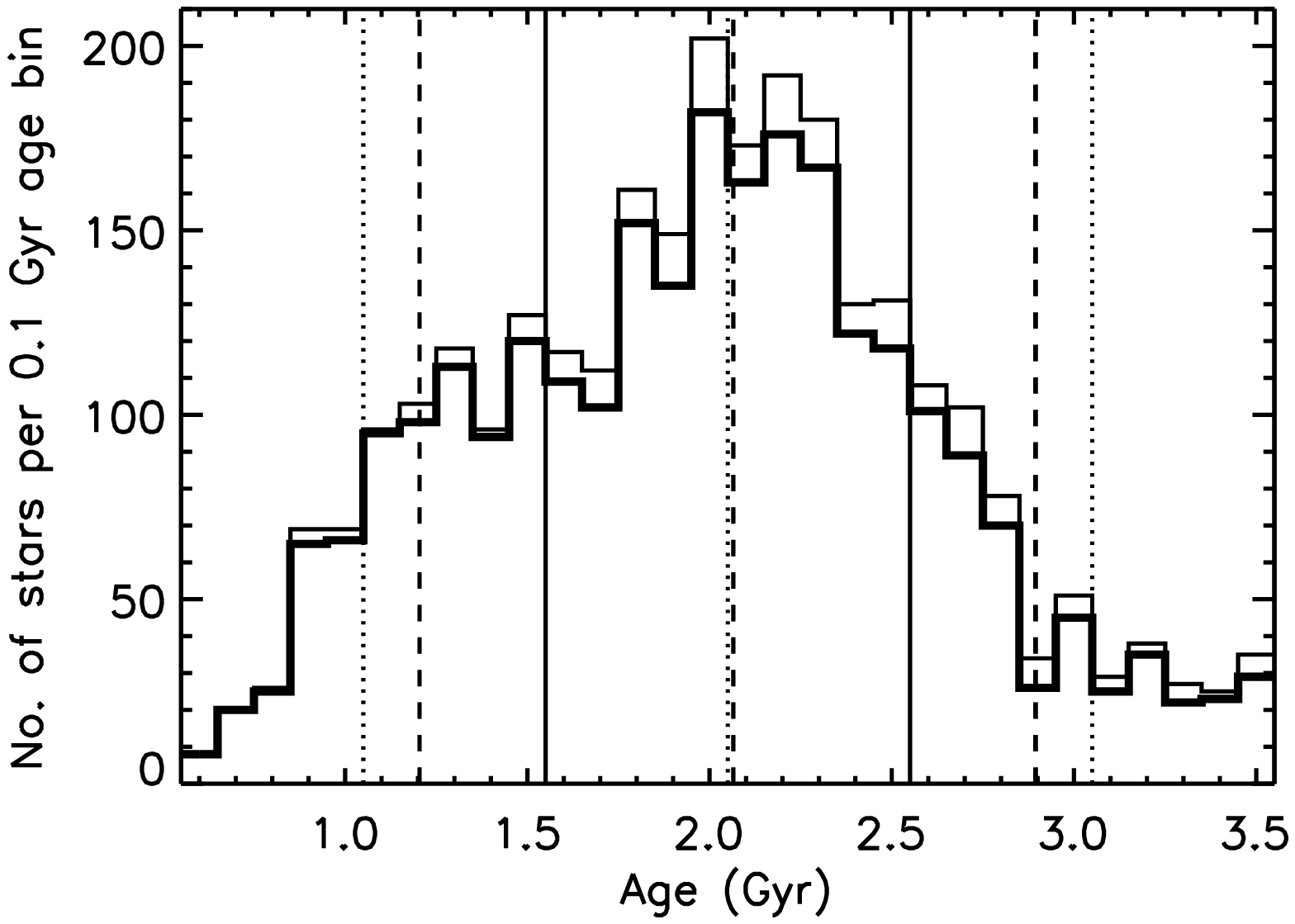,width=0.45\columnwidth}
        \psfig{figure=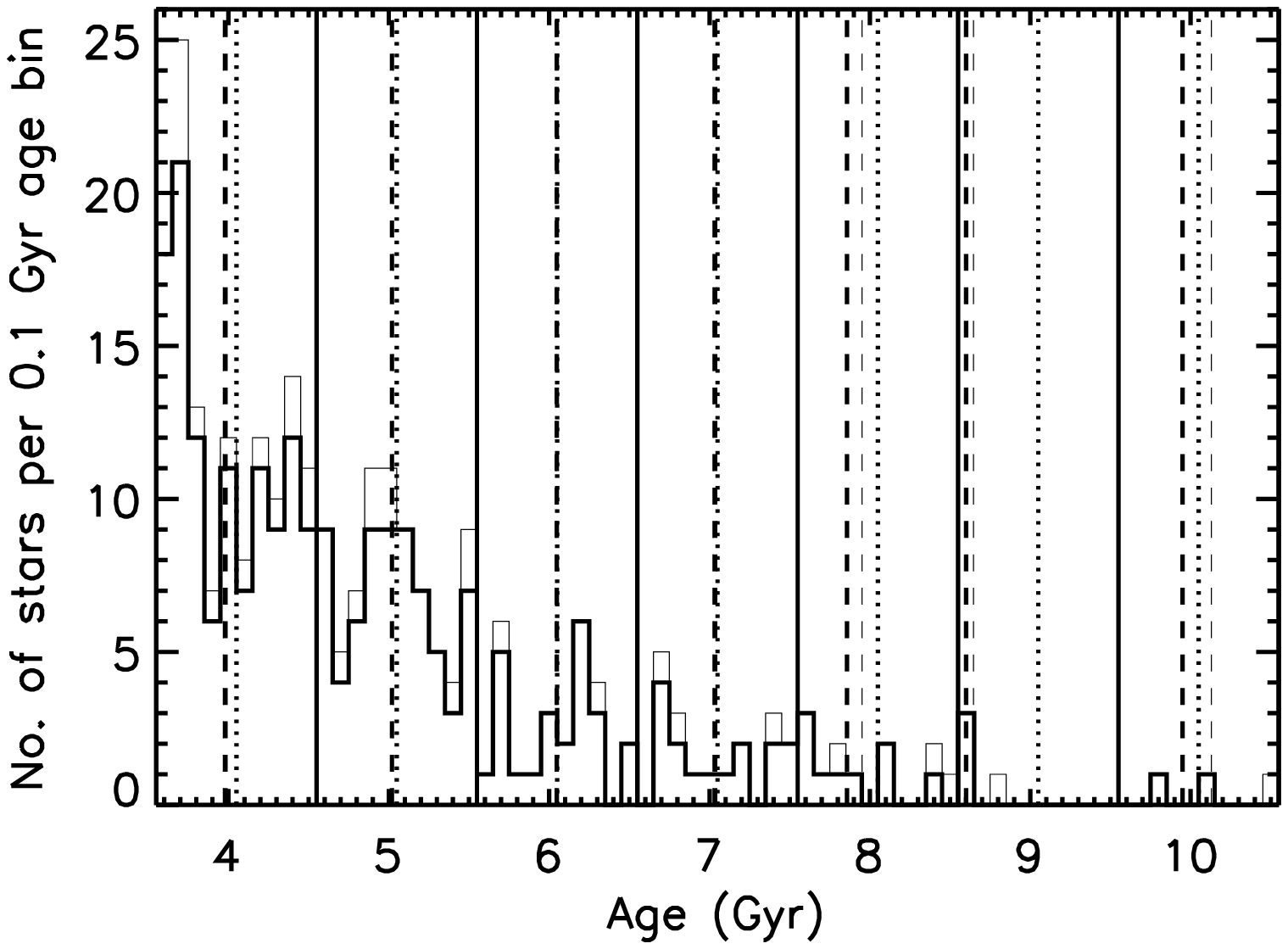,width=0.45\columnwidth} 
\caption{Discrete age distribution of the N04 single stars with absolute age
        errors $<$ 1 Gyr, split between the youngest 3 age bins (left) and the oldest 7 age bins (right).  The thin lines correspond to the whole sample and the thick
        lines are the subsample excluding the Hercules stream.
Because the N04 catalogue stellar ages are provided to the nearest 0.1
        Gyr, each 0.1 Gyr histogram bin only contains stars with the exact age
        that its histogram step is centred on e.g. the youngest histogram bin,
        centred on 0.6 Gyr, only
        contains stars with exactly 0.6 Gyr ages but the histogram step spans 0.55 to 0.65
        Gyr.  The vertical solid lines (including the vertical edges of the
        plots) are the boundaries to the age bins.  
The dotted lines are the central ages of each age bin.  
The robust mean ages
        within each age bin for the whole sample (thin dashed lines) and subsample (thick dashed
        lines) are indisguishable until the three oldest age bins.}
                \label{fig:age1Gyr}
  \end{minipage}              
\end{figure*}

\begin{figure*}
\begin{minipage}{170mm}
\centering
\psfig{figure=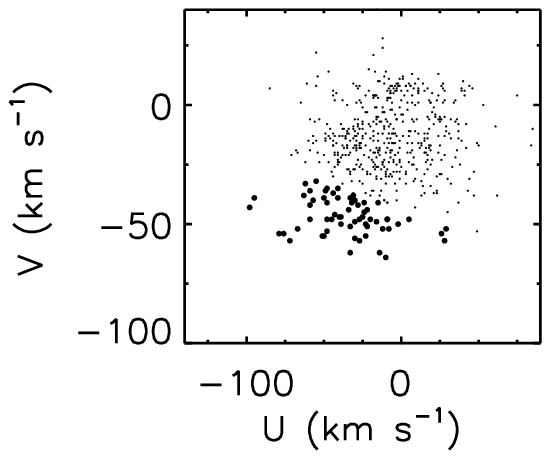} 
\psfig{figure=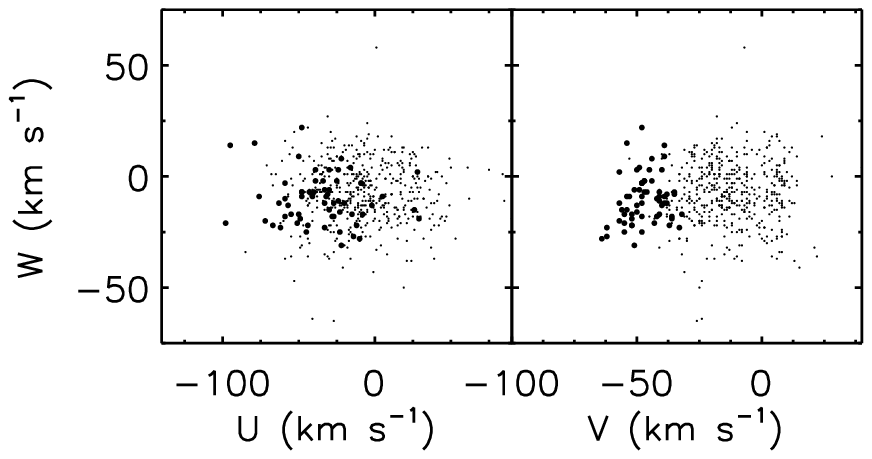}
\caption{$U-V$ (left), $U-W$ (middle) and $V-W$ (right) space velocity diagrams of
  the 527 N04 single stars with 
  absolute age
        errors $<$ 1 Gyr in the 2.5
  $<$ age $\le$ 3.5 Gyr age bin.  We have assigned 62 N04 stars in this age bin to the
        Hercules stream (filled circles) using the Hercules stream
        $UV$ phase-space (defined by F05, see Fig. \ref{fig:uv}) and the Hercules stream
        $W$ phase-space (defined by F05, see Figs. \ref{fig:uw} and
  \ref{fig:vw}).}
          \label{fig:uvw1age}
\end{minipage}
\end{figure*}

\begin{figure*}
\begin{minipage}{170mm}
\centering
 \psfig{figure=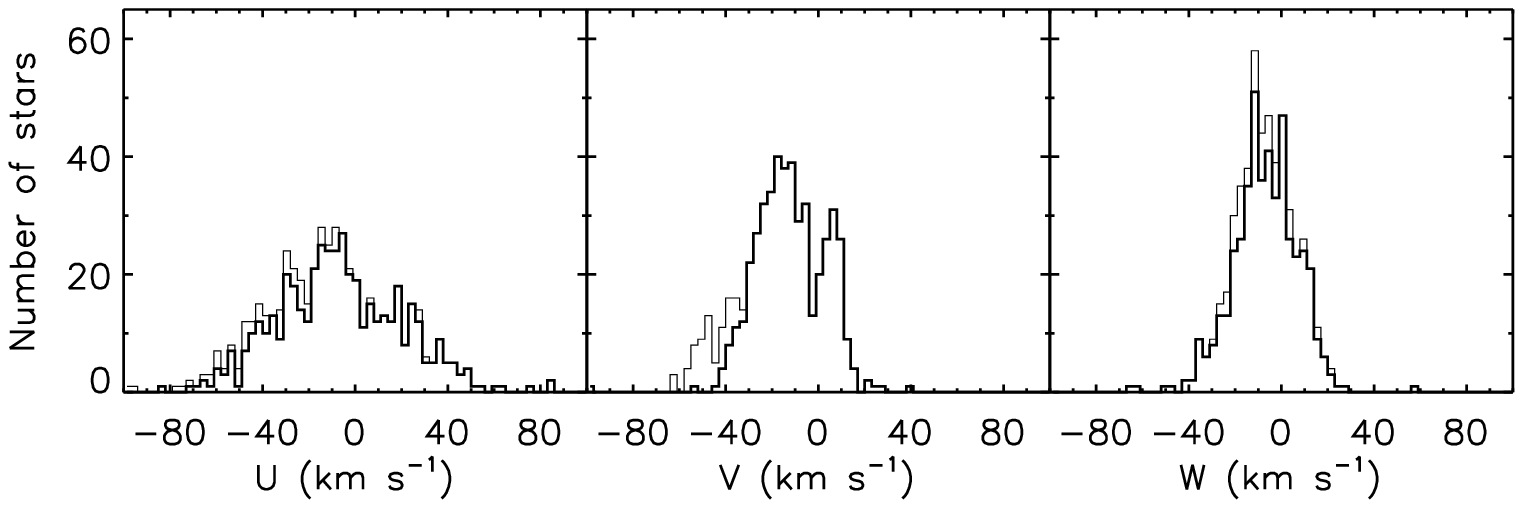}
\caption{$U$ (left), $V$ (middle) and $W$ (right) velocity distributions of
 the N04 single stars with absolute age
        errors $<$ 1 Gyr in the 2.5
  $<$ age $\le$ 3.5 Gyr age bin (thin
  line) and a subsample of these stars that excludes the Hercules stream
  (thick line), as defined in
  Figs. \ref{fig:uv}, \ref{fig:uw} and  \ref{fig:vw}.} 
\label{fig:uvw1hist}
\end{minipage}
\end{figure*}

\begin{figure*}
\centering
\begin{minipage}{170mm}
\includegraphics[width=1.0\columnwidth]{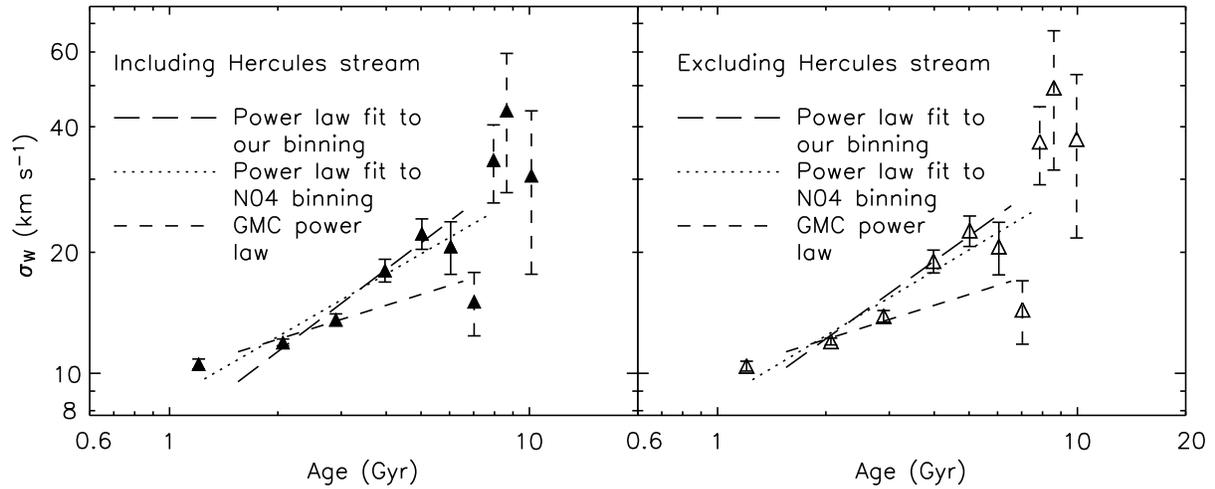}
\caption{Our logarithmic age-$\sigma_{W}$ relation: robust $\sigma_{W}$ of N04 single stars with absolute age
        errors $<$ 1 Gyr in each of the 10 age bins in Fig. \ref{fig:appendixuvw1hist} as a function of the robust mean age of each
        bin including (left, filled triangles) and excluding (right, open triangles) the Hercules
        stream.  As in N04, the youngest age bin has been
         excluded from our power law fit (long dashed line) to avoid biases due to unrelaxed young
         structures.  The oldest age bin included in our fit is centred on 6
        Gyr.  Older age bins are excluded because they contain insufficient numbers
        of stars (see Fig. \ref{fig:appendixuvw1hist}) for their $\sigma_{W}$ values to
        be statistically significant (dashed error bars).  The distribution of $W$ velocities in the age bins centred on
        10 Gyr do not yield robust values of $\sigma_{W}$.  Therefore, their standard
        deviation is plotted instead.}
\label{fig:log1Gyr}
\end{minipage}
\end{figure*}

\begin{figure*}
\begin{minipage}{170mm}
\centering
\includegraphics[width=1.0\columnwidth]{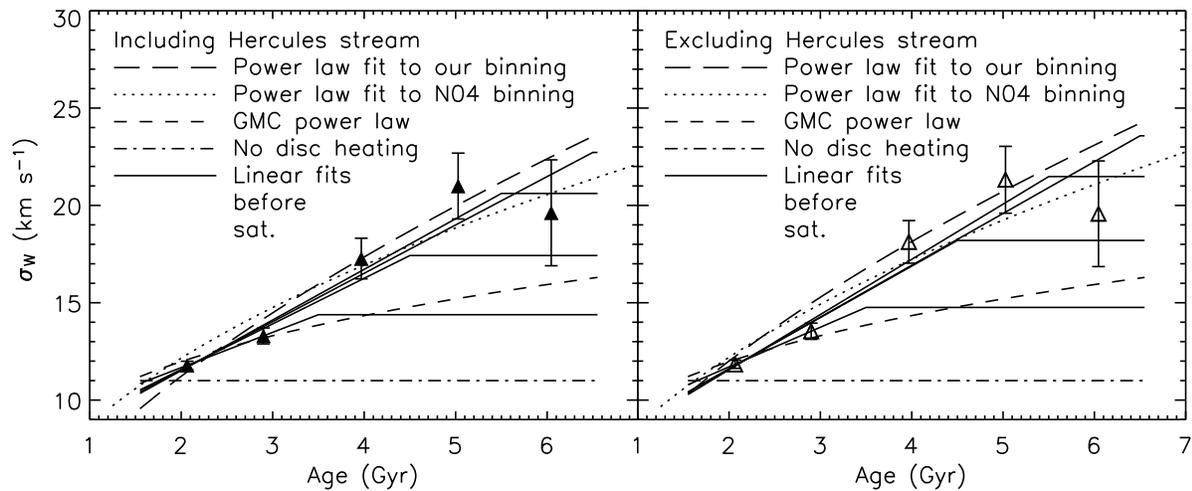}
\caption{Our linear age-$\sigma_{W}$ relation: robust $\sigma_{W}$ of N04 single stars with absolute age
        errors $<$ 1 Gyr in the 5 age bins included in our power law fit in
        Fig. \ref{fig:log1Gyr} as a function of the robust mean age of each
        bin including (left, filled triangles) and excluding (right, open triangles) the Hercules
        stream.  The dashed lines are models representing no disc heating,
        where the constant saturation level is set by the dispersion in the
        youngest bin in Fig. \ref{fig:log1Gyr}.  The solid lines are linear fits to each set of data points up
        to 2.5, 3.5, 4.5, 5.5 and 6.5 Gyr.  Thereafter these models saturate at
        constant dispersions.}  
        \label{fig:linear1Gyr}
\end{minipage}
\end{figure*}

 \begin{table*}
  \centering
  \begin{minipage}{170mm}
 \centering
\label{tab:age1Gyr}
  \caption{$\chi^{2}$ goodness-of-fit test results between our age binning
    data
    and disc heating models in Fig. \ref{fig:linear1Gyr}. 
  $n$ is the number of data points included in the model fit.
$n_{c}$ is the
number of constraints on the model fit. 
 The number of degrees of freedom is $\nu = n - n_{c}$.  The $\chi^{2}$ probability distribution function for $\nu$ degrees of freedom,
$Q(\chi^{2}|\nu)$, is the standard statistical significance of the $\chi^{2}$
test ($Q(\chi^{2}|\nu) = 1 - P(\chi^{2}|\nu)$). If
    $Q(\chi^{2}|\nu) > 0.003$ ($P(\chi^{2}|\nu) < 0.997$), the data accept the model at the
    $3\sigma$ statistical significance level. If $Q(\chi^{2}|\nu) <$ 0.003 ($P(\chi^{2}|\nu) > 0.997$), the data reject the model at the
    $3\sigma$ statistical significance level.} 
  \begin{tabular}{@{}lrrrlrrrl@{}} 
  \hline
Model  & $n$ & $n_{c}$  & $\nu$ & Hercules & $\chi^{2}$ & $\chi^{2}/\nu$ & $Q(\chi^{2}|\nu)$ & $3\sigma$ significance\\
\hline 
Our power law fit & 5 & 2 & 3 & Included & 8.5 & 2.8 & 0.037 & The data accept the model\\
                           &&&& Excluded & 17.4 & 5.8 & 0.001 & The data reject the model\\
\hline 
N04 power law fit & 5 & 2 & 3 & Included & 17.7 & 4.4 & 0.001 & The data reject the model\\
                           &&&& Excluded & 17.0 & 5.7 & 0.001 & The data reject the model\\
\hline
GMC power law & 5 & 2 & 3 & Included & 68.0 & 22.7 & $<10^{-16}$ & The data reject the model\\
                         &&&& Excluded & 28.4 & 9.5 & $10^{-6}$& The data reject the model\\ 
 \hline
No disc heating & 5 & 1 & 4 & Included & 126.1 & 31.5 & $<10^{-16}$ & The data reject the model\\
                         &&&& Excluded & 140.9 & 35.2 & $<10^{-16}$& The data reject the model\\ 
\hline
Linear fit before saturation at 3.5 Gyr & 5 & 3 & 2 & Included & 26.6 & 13.3 & $10^{-6}$ & The data reject the model\\
                                                 &&&& Excluded & 27.2 & 13.6 & $10^{-6}$ & The data reject the model\\
\hline
Linear fit before saturation at 4.5 Gyr & 5 & 3 & 2 & Included & 7.1 & 3.6 & 0.028 & The data accept the model\\
                                                 &&&& Excluded & 6.2 & 3.1 & 0.045 & The data accept the model\\
\hline
Linear fit before saturation at 5.5 Gyr & 5 & 3 & 2 & Included & 3.5 & 1.7 & 0.175 & The data accept the model\\
                                                 &&&& Excluded & 3.9 & 1.9 & 0.146 & The data accept the model\\
\hline
Linear fit before saturation at 6.5 Gyr & 5 & 3 & 2 & Included & 4.0 & 2.0 & 0.136 & The data accept the model\\
                                                 &&&& Excluded & 4.7 & 2.3 &  0.097 & The data accept the model\\
\hline
\end{tabular}
\end{minipage}
\end{table*}

Higher resolution age binning is required to try to further rule out other disc heating models.  Linear age binning, which needs bins of constant age range, will sample the old thin disc with more data points.  Selecting the range
of each age bin to be 1 Gyr wide more than doubles the number of data
points sampling the old thin disc and thick disc.  It also more than
halves the number of age bins sampling the young thin disc but this
regime has already been well constrained in the previous section.  

We also chose an absolute age error of $<$ 1 Gyr so that any
star can only shift laterally by a maximum of one age bin.  This is not the case with the N04 binning.  The preponderance of young stars in the N04 magnitude-limited sample (because of their higher luminosity) means their bins are more narrow than the typical age errors ($<$ 25 per cent) so the errors of some young stars span more than one age bin.  Absolute
age errors $<$ 1 Gyr means selecting stars with
$\sigma_{\mathrm{age}}^{\mathrm{high}}$ - age $<$ 1 Gyr and age -
$\sigma_{\mathrm{age}}^{\mathrm{low}}$ $<$ 1 Gyr.  Selecting single
stars with absolute age errors $<$ 1 Gyr actually samples more stars
(3,083) than the relative age error of $<$ 25 per cent (2,801) because
the vast majority of N04 stars are between 1 and 3 Gyr old (see N04's
fig. 17).   The youngest and oldest stars with an absolute age error $<$ 1 Gyr are 0.6 and 10.5 Gyr,
constraining our age interval to 0.5 $<$ age $\le$ 10.5 Gyr, giving us
ten 1 Gyr wide age bins.  Fig. \ref{fig:age1Gyr} shows the age distribution of our new sample.

Figs. \ref{fig:uvw1age} and \ref{fig:uvw1hist} show the space velocity
diagrams and histograms respectively of a representative age bin from our new
sample (plots of all our age bins are in the Appendix) that is most similar in age to
the N04 age bin in Figs. \ref{fig:uvw25age} and \ref{fig:uvw25hist}.  Unlike
the N04 binning in Fig. \ref{fig:uvw25hist} (and
Fig. \ref{fig:appendixuvw25hist}), the age bin in Fig. \ref{fig:uvw1hist} (and
the two younger bins, all representing the young thin disc, in
Fig. \ref{fig:appendixuvw1hist}) do not suffer from statistical noise.  They
show the non-Gaussianity of the $U$ and $V$ distributions compared to $W$ more
clearly than in Figs. \ref{fig:uvw25hist} and \ref{fig:appendixuvw25hist}.  Again, the young thin disc $W$ histograms do not look completely relaxed.  Fig. \ref{fig:appendixuvw1hist} clearly shows that
only including stars with errors $<$ 1 Gyr severely reduces the numbers of
stars in the older bins.  The oldest $W$
distribution that is still approximately a complete symmetric Gaussian and statistically significant with 26 stars is in
bin 5.5 $<$ age $\le$ 6.5 Gyr.  Therefore, after the removal of statistically insignificant age bins older than 6.5 Gyr, our higher age resolution binning only adds one extra bin with which to constrain the age-$\sigma_{W}$ relation of the old thin disc. The 5.5 $<$ age $\le$ 6.5 Gyr is the oldest age bin included in
the power law fit in Fig. \ref{fig:log1Gyr}.  The power law fit to our binning produces a higher scaling (heating) 
exponent than the N04 binning power law: $k$ = 0.63 $\pm$ 1.21 including the Hercules stream ($k$ = 0.60
$\pm$ 1.22 excluding the Hercules stream).

We plot all the age bins from Fig. \ref{fig:age1Gyr} in Fig. \ref{fig:log1Gyr} but differentiate between the
statistically significant and insignificant bins.  Although not included in
the power law fit, the oldest age bin in N04's binning in Fig. \ref{fig:log25}
(and N04's fig. 31), seems to be a continuation of the power law, suggesting
it is part of a continuing heating process.  The oldest age bin includes some
thick disc stars, which have been shown by many authors to have approximately
double the thin disc $\sigma_{W}$.  This thick disc signature is not seen in
the N04 binning because the oldest age bin spans the oldest part of the thin
disc and all the thick disc so it is numerically dominated by old thin disc
stars that decrease $\sigma_{W}$.  Our higher age resolution bins better
resolve the thin disc-thick disc transition region.  The age bins centred on 8
and 9 Gyr in Fig. \ref{fig:log1Gyr} show a modest increase in $\sigma_{W}$
compared to the power law fits, reaching values more associated with the thick
disc, $35 \pm 3$ km s$^{-1}$ \citep{chiba2000}, $39 \pm 4$ km s$^{-1}$
\citep{soubiran2003}, than the thin disc.  This is in agreement with the jumps
in the age-velocity dispersion relation at 8 and 9 Gyr found respectively by
\citet{sommer-larsen1993} and \citet{quillen2001}, the latter being
significant at the 2-3$\sigma$ level.  The abrupt increase is interpreted as a
formation signature of the thick disc, which is consistent with models where
the thick disc was formed by dramatic disc heating when a satellite galaxy
falls onto the Galactic disc in a minor merger $\sim$ 10 Gyr ago \citep{gilmore1989,freeman1991,toth1992,quinn1993}.  Despite their statistical insignificance, it is tempting to associate these bins with the formation of the thick disc.

Again as expected, Figs. \ref{fig:log1Gyr} and \ref{fig:linear1Gyr} show that
the largest effect in removing the Hercules stream from our binning sample is
seen in the oldest thick disc age bins, which are not included in the
age-$\sigma_{W}$ relation.  Figs. \ref{fig:log1Gyr}, \ref{fig:linear1Gyr} and
Table 4 show that removing the Hercules stream has a negligible effect on the
age-$\sigma_{W}$ relation, allowing the relation to be constrained by our age
binning as well as the N04 binning.  

The same disc heating models as before are considered with the results in
Table 4.  Again, we have plotted the GMC power law
in Figs. \ref{fig:log1Gyr} and \ref{fig:linear1Gyr}, setting $a = \sigma_{0}$
= 10 km s$^{-1}$ so that at the youngest age included in the fits (1.55 Gyr),
$\sigma_{W} \approx$ 11.2 km s$^{-1}$.  With our binning there are fewer degrees of freedom, so the
$Q(\chi^{2}|\nu)$ values are generally smaller in Table 4 than in Table 3 for
the N04 binning.  Our binned data including the Hercules stream accept our
power law model but marginally reject it when the stream is excluded.  Both
including and excluding the Hercules stream in our binned data also marginally
reject the N04 power law fit, strongly reject a no disc heating model, reject
the GMC power law (although as discussed earlier, this may not be conclusive),
reject a linear fit before saturation at 3.5 Gyr model but accept models with
linear fits before saturation at 4.5,  5.5 and 6.5 Gyr.  Hence, our binning
reinforces results from the previous section that the data do not require a
power law and actually statistically prefer saturation at $\sim$ 4.5 Gyr. 

\section{Discussion}
\label{s:discussion}

This paper re-visits the Galactic thin disc age-velocity dispersion relation.  We show that dynamical streams complicate the $U$ and $V$ velocity distribution functions such that  a straight-forward dispersion is not an adequate parametrization of their distribution functions.  This calls into question the appropriateness of power law fits to the $\sigma_{U}$ and $\sigma_{V}$ relations with stellar age.
  It shows that observationally measured $\sigma_{U}$ and $\sigma_{V}$ can only be used to constrain in-plane disc heating models if the N04 sample can be kinematically decomposed (like in F05) so that only the smooth velocity ellipsoidal background is used, excluding the dynamical streams.  Their strong presence in the solar neighbourhood may cast  doubt on the applications of $\sigma_{U}$ in determining the asymmetric drift, the extrapolating to zero $\sigma_{U}$ to recover the $V$ solar motion and the interpretation of the velocity ellipsoid dispersion axis ratios $\sigma_{V}$/$\sigma_{U}$, $\sigma_{W}$/$\sigma_{U}$ and $\sigma_{W}$/$\sigma_{V}$.  F05 have already pointed out using their sample that the underlying velocity ellipsoid, after the removal of streams is not centred on the commonly accepted radial anti-centre motion.  Instead it is centred on $\langle U \rangle = -2.78 \pm 1.07$ km s$^{-1}$.  Inclusion of the streams yields the commonly accepted value: $\langle U \rangle = -10.25 \pm 0.15$ km s$^{-1}$. 

The age-$\sigma_{W}$ relation can be constrained using the N04 data set because the dynamical streams are well phase-mixed in $W$, as illustrated by removal of the Hercules stream having no effect on the relation, which also suggests that the Galactic bar does not vertically heat the local disc.  We do not find any signature of the stellar warp in the Galactic disc so we do not need to exclude any N04 stars from the age-$\sigma_{W}$ relation on this basis, because they are all free from this potential source of $\sigma_{W}$ bias.

We reproduce the N04 age-$\sigma_{W}$ relation and confirm their finding that
vertical disc heating does not saturate at an early stage (before $\sim$ 4.5
Gyr).  Their logarithmic age binning naturally led them to fit a power law.
However, they did not consider other disc heating models and so their
conclusion was that the power law fit was evidence that vertical disc heating
is continuous throughout the lifetime of the thin disc.  We have considered
other disc heating models that saturate after different epochs.  Our key new
result is that a power law is not required by their data.  A power law fit is
statistically similar to disc heating models which saturate after $\sim$ 4.5
Gyr (including and excluding the Hercules stream).  These saturation model
fits are consistent with the observational turnover at $\sim$ 5 Gyr found by \citet{carlberg1985b}.  Simulations of vertical disc heating solely from GMCs find $k$ = 0.26 \citep{hanninen2002}.  Figs. \ref{fig:linear25} and \ref{fig:linear1Gyr} show that this power law is consistent with a minimal increase in $\sigma_{W}$ for old stars, not too dissimilar to the models that saturate after 3.5 and 4.5 Gyr.  

To see if the data could rule out any of the disc heating models considered, we examined the relation at a higher age resolution but found very similar results to the N04 age binning.  Therefore, two diametrically different age binning methods using the best data available for constraining the age-$\sigma_{W}$ relation still cannot distinguish between competing disc heating models.  The issue is unlikely to be resolved until there are many more space velocities and ages measured for old thin disc stars.  This may need to wait until the ESA {\it Gaia} satellite era \citep{perryman2001}. 

Nevertheless, our results provide new constraints to be fulfilled by disc heating simulations.  Prior to this paper, new disc heating simulations would have tried to explain the continued disc heating found in all three cardinal directions by N04.  Now, we have shown that new simulations cannot use in-plane age-velocity dispersion relations derived using stellar samples containing dynamical streams and straight-forward dispersions.  Also, the new simulations must be able to reproduce the complex $U$ and $V$ substructure in the solar neighbourhood, caused by dynamical streams, while modelling in-plane disc heating mechanisms (e.g. \citealt{desimone2004}).   They must also take into account that these dynamical streams dominate the solar neighbourhood velocity distribution function and include evaporating star clusters, which may have been created by the spiral wave that spawned the streams. 

Assuming our choice of initial $\sigma_{W}$ and the modelling of \citet{hanninen2002} are both realistic, simulations of vertical disc heating solely from GMCs appear unable to reproduce the amplitude of our age-$\sigma_{W}$ relation on their own.  This suggests that another vertical disc heating mechanism is or has been involved in the evolution of the Galactic thin disc.  We have shown that the data do not rule out heating by halo black holes.  However, this option has no observational support.  Black holes passing through the disc would probably reveal themselves as high proper motion X-ray sources and no such objects have been found. 

In the context of the hierarchical galaxy formation paradigm, another vertical
heating mechanism is minor mergers with dwarf galaxies.  We find extremely
tentative evidence of an abrupt feature in the age-$\sigma_{W}$ relation at
$\sim$ 8 Gyr, which could be the sudden vertical heating of an early thin disc
into the thick disc due to a major merger \citep{quinn1993}.   The relation
does not exhibit any other strong features like this, suggesting it is
unlikely that the Milky Way suffered another major merger in the last $\sim$ 8
Gyr but it does not rule out minor mergers during this period of relatively
quiescent disc evolution.  In Figs. \ref{fig:linear25} and
\ref{fig:linear1Gyr}, the old thin disc bins with $\sigma_{W}$ greater than
the GMC power law may be the signature of minor mergers that took place
$\gtrsim$ 3 Gyr ago at redshift $\gtrsim$ 0.5.  \citet{velazquez1999} showed
that disc galaxies can accrete quite massive satellites without destroying the
disc, particularly if the orbits are retrograde.  They found that a satellite
on a retrograde orbit with an initial mass of 20 per cent of the Milky Way's
disc could increase the velocity ellipsoid at the solar neighbourhood by
$\Delta\sigma_{W}$ = 6 km s$^{-1}$.  The same satellite, but on a prograde
orbit, leads to $\Delta\sigma_{W}$ = 12 km s$^{-1}$.  It is intriguing that
the difference between the observed age-$\sigma_{W}$ relation and the
predicted relation only heated by GMCs is $\sim$ 6 km s$^{-1}$ at an epoch
$\sim$ 5 Gyr ago in Fig. \ref{fig:linear1Gyr}.  The agreement between the
observed age-$\sigma_{W}$ relation and the predicted relation only heated by
GMCs is good between the present day and $\sim$ 3 Gyr ago, which may suggest
that no minor mergers with significant satellite mass have occurred in the
last $\sim$ 3 Gyr.  

The RAdial Velocity Experiment (RAVE, \citealt{steinmetz2006paper}) is measuring the radial velocities of stars further from the Sun than the N04 and F05 samples.  RAVE vertical velocities can be used to test the vertical smoothness of the velocity distribution further from the Galactic plane than possible with N04 and F05, as well as to search for tidal streams falling onto the Milky Way disc from satellite galaxies (Seabroke et al. 2007, in prep.), which may be responsible for some of the vertical disc heating observed in the age-$\sigma_{W}$ relation in the solar neighbourhood.

\section*{Acknowledgments}

GMS gratefully acknowledges financial support from a Particle Physics and Astronomy Research Council PhD
Studentship, the Gordon Wigan Fund, Gonville and Caius College and the Cambridge Philosophical Society Research Studentship Fund.

\bibliographystyle{mn2e}
\bibliography{references}

\appendix

\section[]{Space velocity distributions in all the age bins}
\label{s:appendix}

\subsection[]{N04 age bins}
\label{s:appendixN04}

\begin{figure*}
\begin{minipage}{170mm}
\centering
\psfig{figure=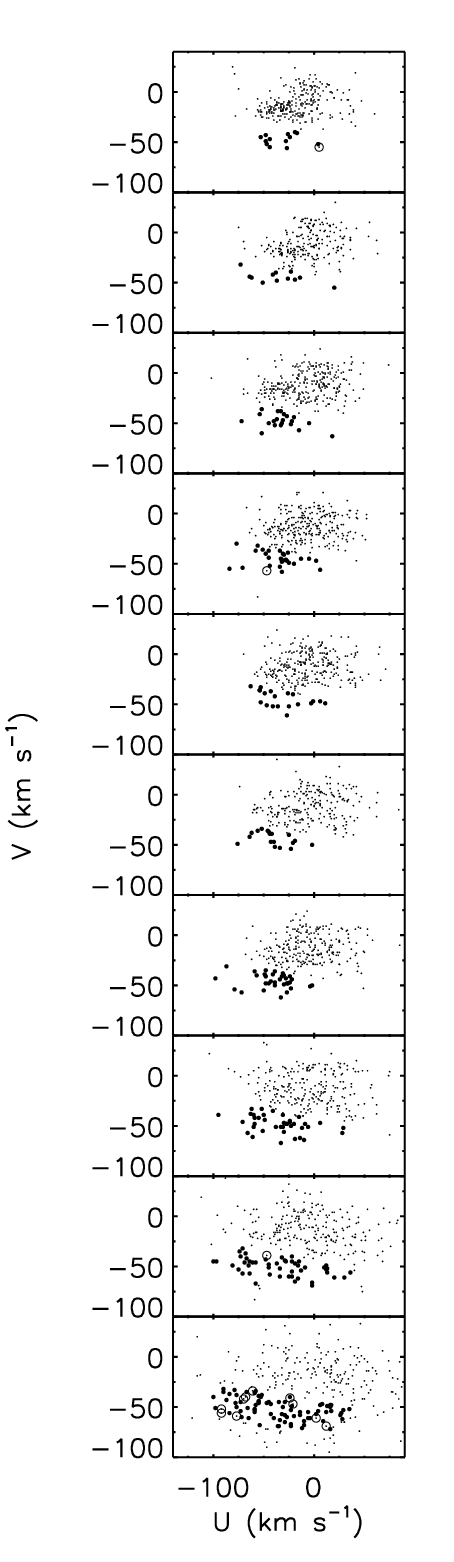} 
\psfig{figure=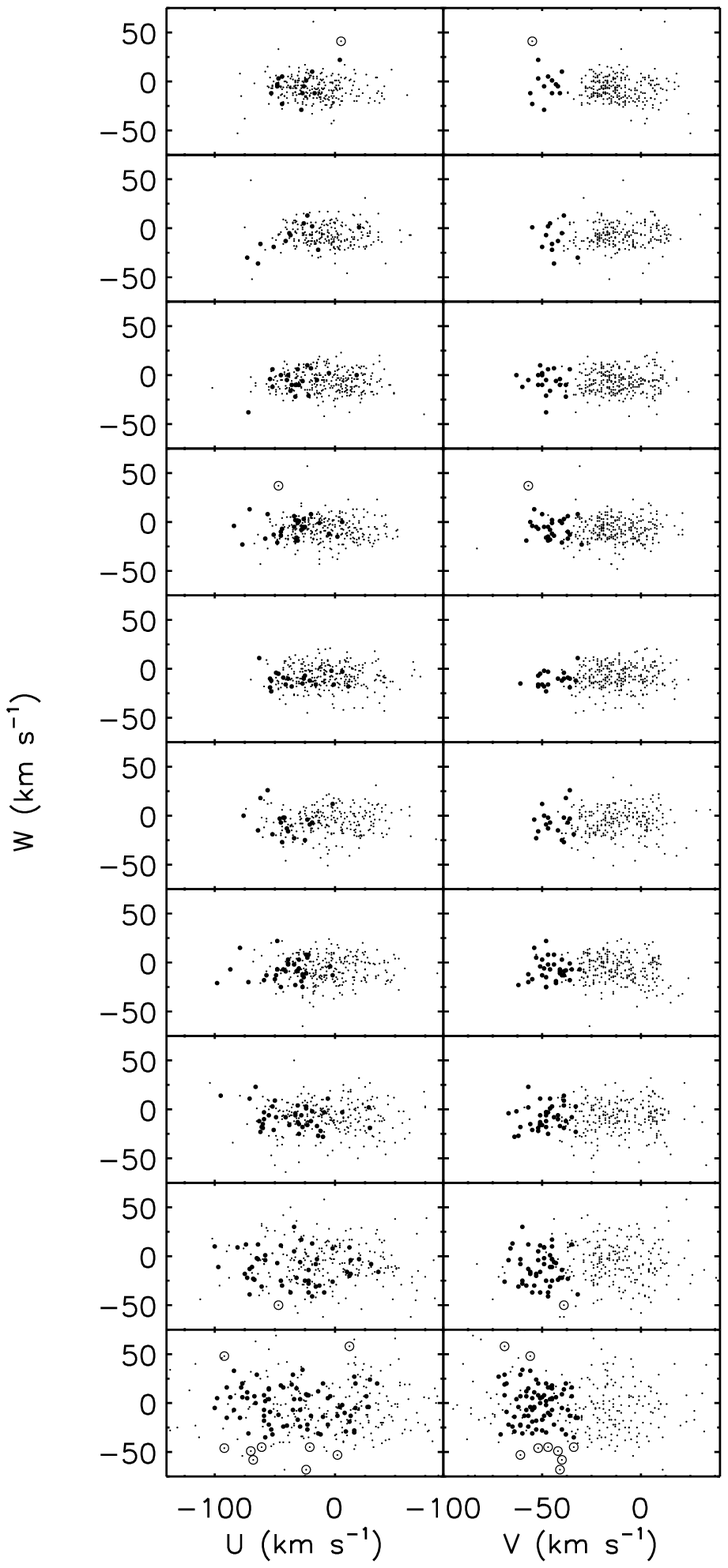}
\caption{$U-V$ (left), $U-W$ (middle) and $V-W$ (right) space velocity diagrams of
  the N04 single stars with relative age errors $<$ 25 per cent plotted in
  the 10 age bins from Fig. \ref{fig:appendixuvw25hist} (youngest at the top and
  oldest at the bottom).  We have assigned N04 stars to the
        Hercules stream (filled circles) and high-velocity stars (open circles) in the Hercules stream
        $UV$ phase-space (defined by F05, see Fig. \ref{fig:uv}) but outside the Hercules stream
        $W$ phase-space (defined by F05, see Figs. \ref{fig:uw} and \ref{fig:vw}).}
\label{fig:appendixuvw25age}
\end{minipage}
\end{figure*}

\begin{figure*}
\begin{minipage}{170mm}
\centering
 \psfig{figure=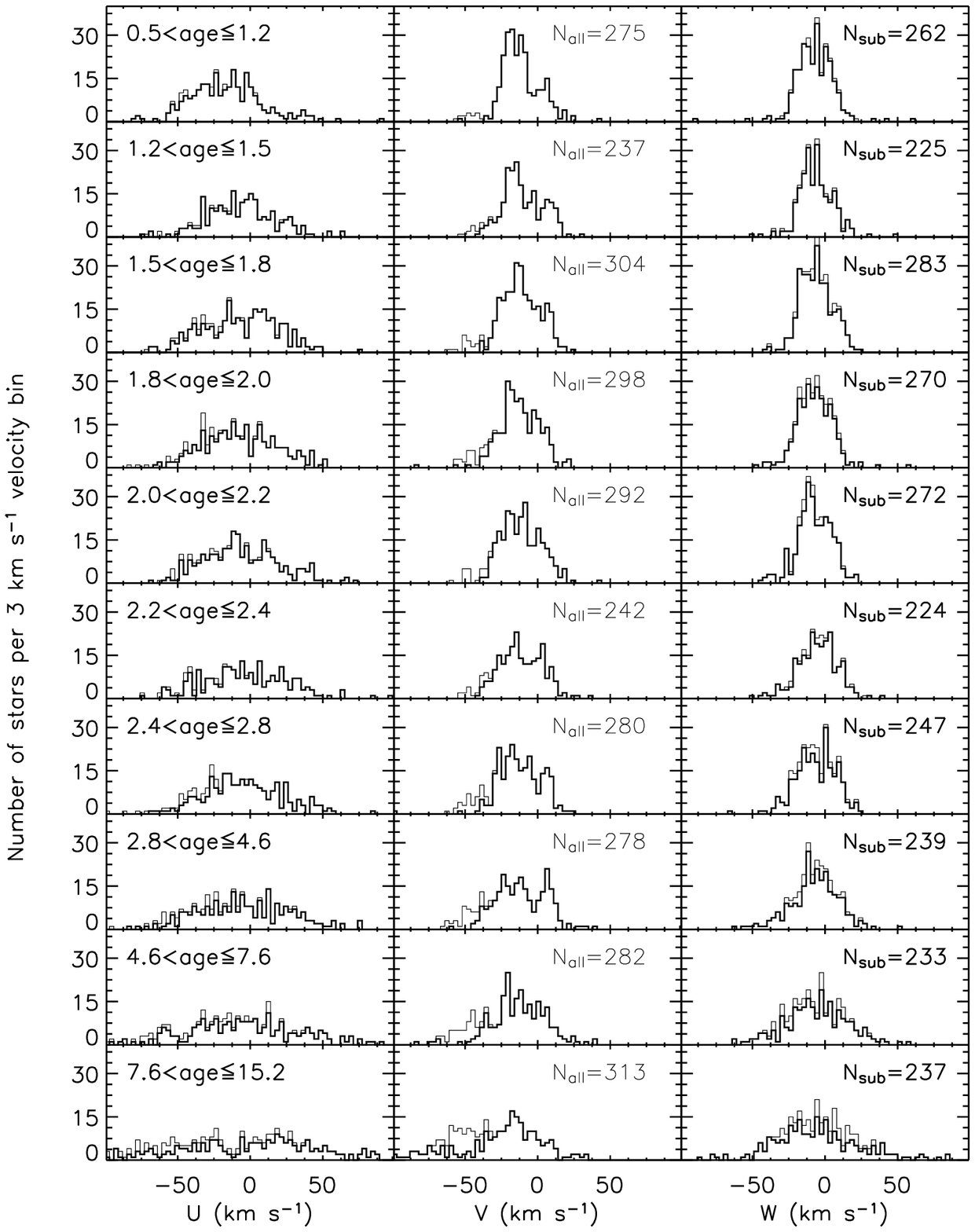}
\caption{$U$ (left), $V$ (middle) and $W$ (right) velocity distributions of the N04 single stars with relative age errors $<$ 25 per cent plotted in
  the 10 age bins that dissect their corresponding age distribution to give
  approximately equal numbers (N$_{\textrm{all}}$) of stars in each age bin (thin
  lines) and a subsample (N$_{\textrm{sub}}$) of these stars that excludes the Hercules stream (as defined in
  Figs. \ref{fig:uv}, \ref{fig:uw} and  \ref{fig:vw}) plotted in the same age
  bins (thick lines). The boundaries
  of each age bin are given in Gyr. }
\label{fig:appendixuvw25hist}
\end{minipage}
\end{figure*}

For completeness, all the N04 age bins have their space velocity diagrams plotted in
Fig. \ref{fig:appendixuvw25age} and their space velocity histograms plotted in
Fig. \ref{fig:appendixuvw25hist}.

\subsection[]{Our age bins}
\label{s:appendix}

\begin{figure*}
\begin{minipage}{170mm}
\centering
\psfig{figure=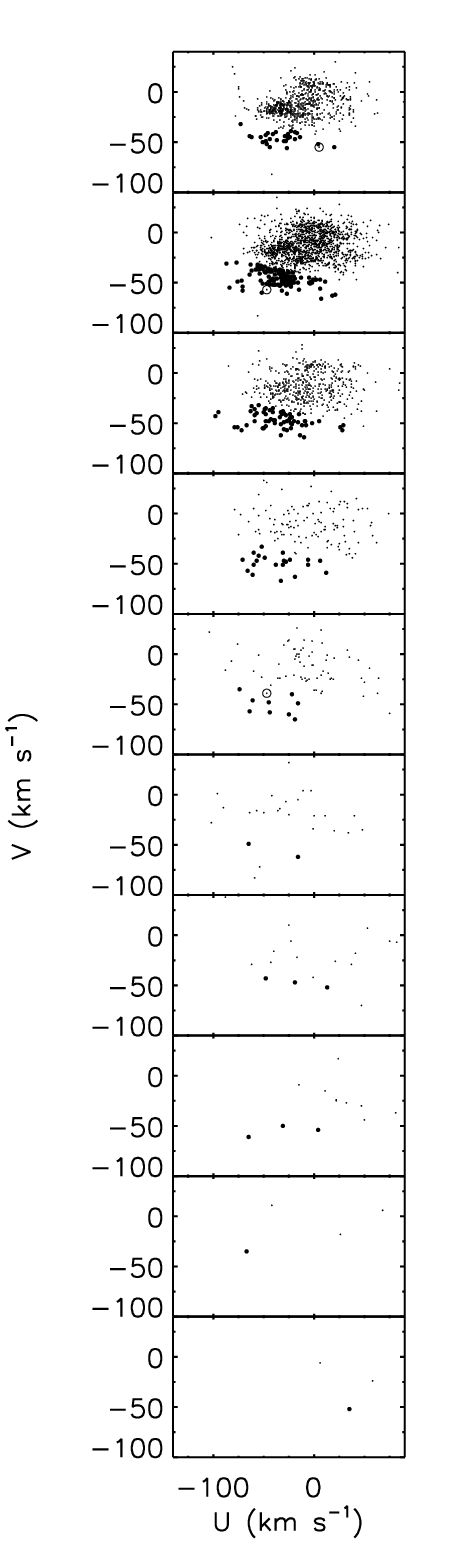} 
\psfig{figure=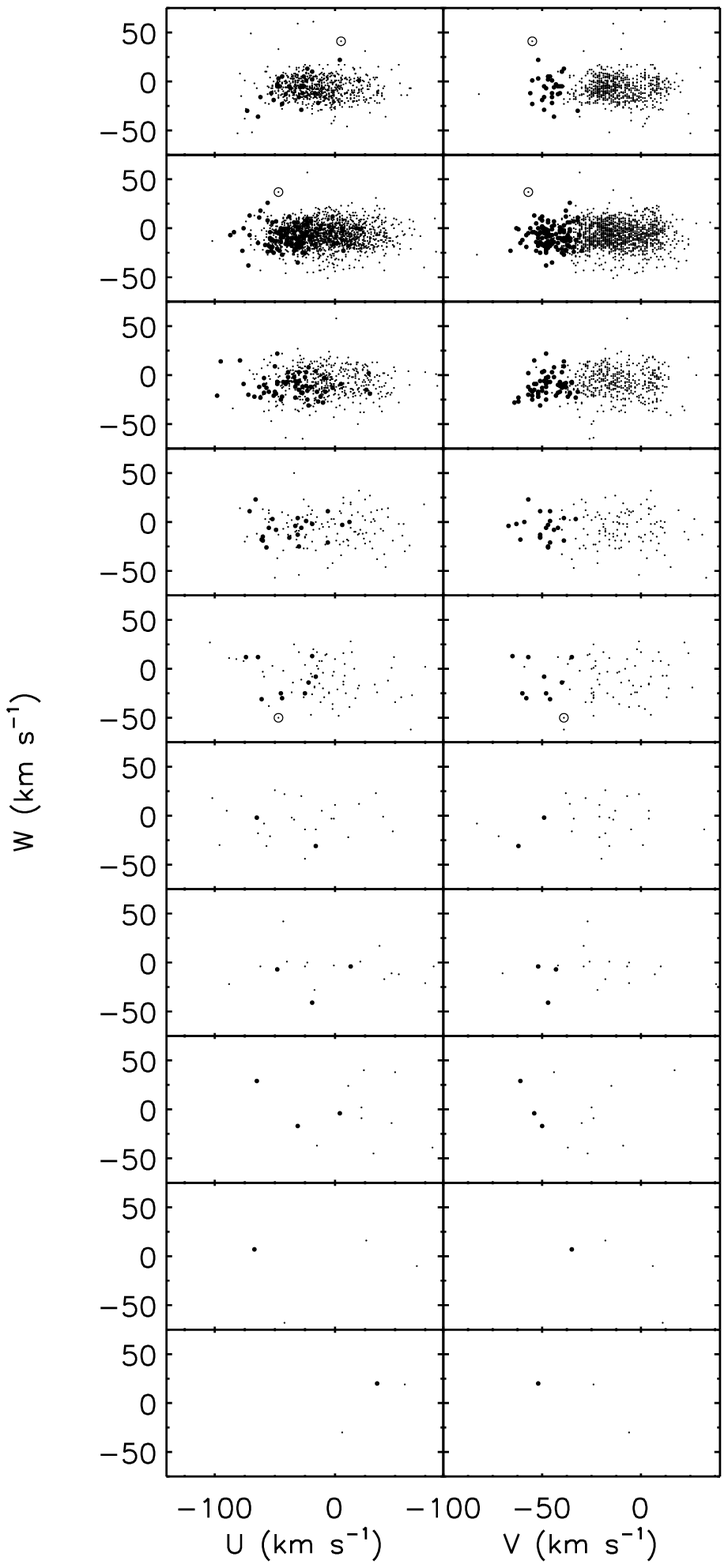}
\caption{$U-V$ (left), $U-W$ (middle) and $V-W$ (right) space velocity diagrams of
  the N04 single stars with 
  absolute age
        errors $<$ 1 Gyr plotted in
  the 10 age bins from Fig. \ref{fig:uvw1hist}
   (youngest at the top and
  oldest at the bottom).  We have assigned N04 stars to the
        Hercules stream (filled circles) and high-velocity stars (open circles) in the Hercules stream
        $UV$ phase-space (defined by F05, see Fig. \ref{fig:uv}) but outside the Hercules stream
        $W$ phase-space (defined by F05, see Figs. \ref{fig:uw} and \ref{fig:vw}).}
          \label{fig:appendixuvw1age}
\end{minipage}
\end{figure*}

\begin{figure*}
\begin{minipage}{170mm}
\centering
 \psfig{figure=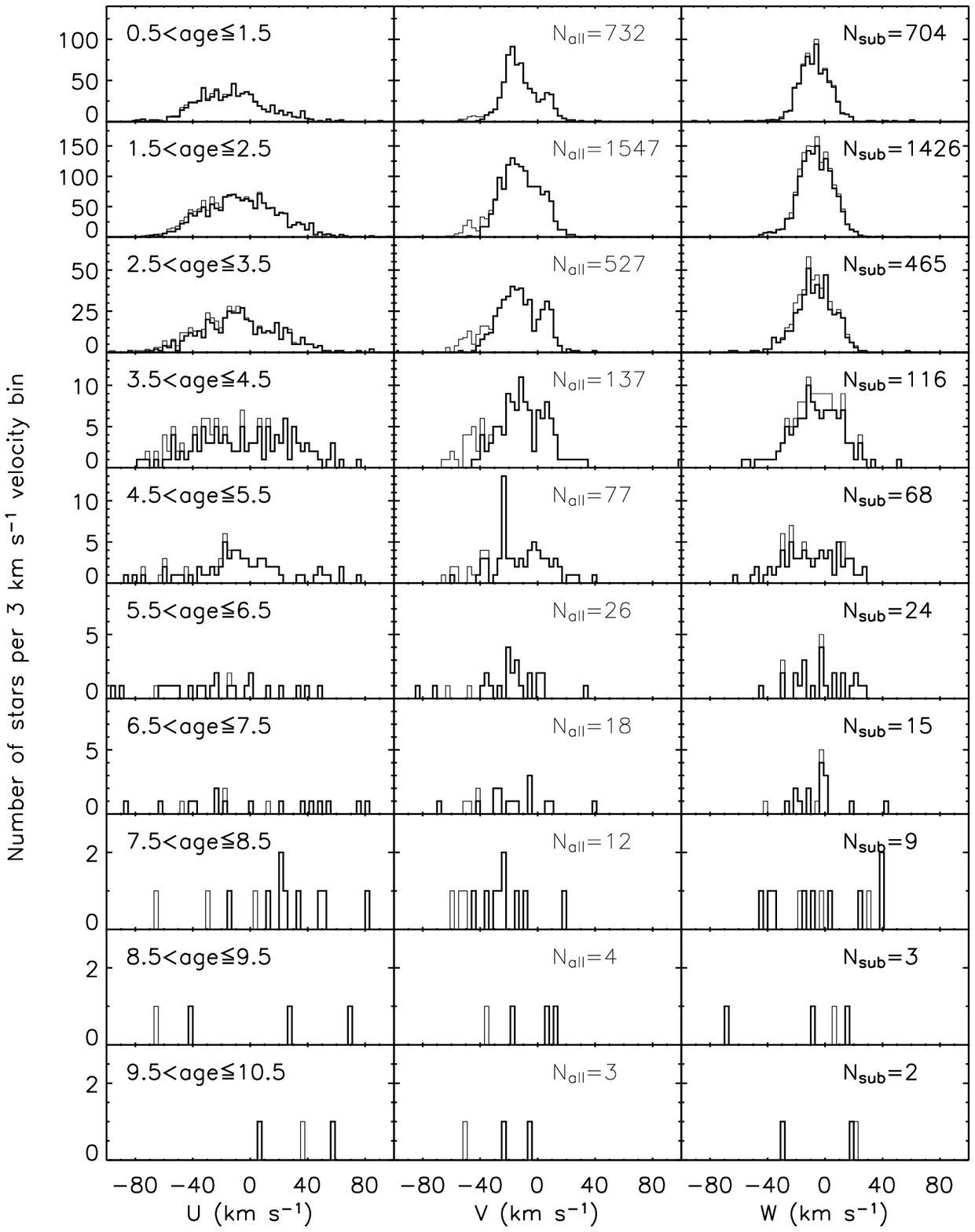}
\caption{$U$ (left), $V$ (middle) and $W$ (right) velocity distributions of the N04 single stars with absolute age
        errors $<$ 1 Gyr plotted in
  the 10 age bins that dissect their corresponding age distribution to give
        equal width (1 Gyr) age bins.  The boundaries
  of each age bin are given in Gyr.  N$_{\textrm{all}}$ is the total number of stars in each age bin (thin
  lines) and N$_{\textrm{sub}}$ is the number of these stars in the subsample that excludes the Hercules stream (as defined in
  Figs. \ref{fig:uv}, \ref{fig:uw} and  \ref{fig:vw}) plotted in the same age
  bins (thick lines).}
\label{fig:appendixuvw1hist}
\end{minipage}
\end{figure*}

For completeness, all our age bins have their space velocity diagrams plotted in
Fig. \ref{fig:appendixuvw1age} and their space velocity histograms plotted in
Fig. \ref{fig:appendixuvw1hist}.

\bsp

\label{lastpage}

\end{document}